\documentclass[journal=jacsat,manuscript=article]{achemso}
\SectionNumbersOn

\usepackage{color,soul}
\usepackage{xspace}
\usepackage{xcolor}
\usepackage{tablefootnote}

\usepackage[version=3]{mhchem} 
\usepackage{mathptmx}
\usepackage{booktabs}

\usepackage{siunitx}
\usepackage{physics}
\usepackage{mhchem}

\author{Mikhail A. Anisimov}%
\email{anisimov@umd.edu}
\affiliation{Institute for Physical Science and Technology, University of Maryland, College Park, MD 20742, USA}
\alsoaffiliation{Department of Chemical and Biomolecular Engineering, University of Maryland, College Park, MD 20742, USA}

\author{Sergey V. Buldyrev}%
\affiliation{ Department of Physics, Yeshiva University, New York, NY 10033, USA}
\alsoaffiliation{Department of Physics, Boston University, MA 02215, USA}

\author{Fr\'ed\'eric Caupin}
\affiliation{Institut Lumi\`ere Mati\`ere, Universit\'e Claude Bernard Lyon 1, CNRS, Institut Universitaire de France, F-69622 Villeurbanne, France}

\author{Thomas J. Longo}%
\affiliation{Institute for Physical Science and Technology, University of Maryland, College Park, MD 20742, USA}

\title{Degenerate Fluid Polyamorphism Induced by Symmetrical Molecular Interconversion}

\abbreviations{IR,NMR,UV}
\keywords{American Chemical Society, \LaTeX}

\begin{document}








\begin{abstract}
Fluid polyamorphism is the existence of multiple fluid-fluid phase transitions in a single-component substance. It can occur due to interconversion between two alternative molecular or supramolecular states. In this work, we investigate a special (``degenerate'') case of fluid polyamorphism, in which all three characteristic parameters of the interconversion equilibrium constant, i.e. the changes of energy, entropy, and volume, are zero. This feature of interconversion is typical for the Ising spin model of ferromagnets but is also observed in a polyamorphic chiral fluid mixture of interconverting enantiomers. To investigate the consequences of interconversion's degeneration for fluid polyamorphism, we have performed a meanfield analysis and 3D Monte Carlo simulations of a compressible binary lattice with interconverting species (referred to as the ``blinking-checkers model''), which generally demonstrates the existence of both liquid-gas and liquid-liquid transitions. By tuning the interaction parameters, we have demonstrated that, in the degenerate interconverting case, a coupling between the fraction of interconversion (a vector-like nonconserved order parameter) and the total density (a scalar conserved order parameter) may produce a symmetrical tricritical point. At this point, the line of second-order transitions between two fluids, ``disordered'' (with 50:50 interconversion) and ``ordered'' (with temperature and pressure dependent interconversion fraction), is terminated by first-order transitions between the fluid states. This point exhibits the typical features of symmetrical tricritical points as observed in a superfluid mixture of helium isotopes and in some magnetic materials. We also show that the transition between the ordered and disordered fluid could occur in either the liquid or the gaseous phase. 
\end{abstract}

\section{Introduction}~\label{Sec_Introduction}
The phenomenon of fluid polyamorphism, the existence of multiple fluid-fluid phase transitions in a single-component substance~\cite{Debenedetti_Water_1998,Stanley_Liquid_2013,Tanaka_Liquid_2020}, can phenomenologically be described by the emergence of an additional order parameter, the equilibrium fraction of interconversion between alternative molecular or supramolecular states~\cite{Anisimov_Polyamorphism_2018}. The fraction of interconversion (a nonconserved property) may be coupled with another order parameter, associated with the density (a conserved property), producing various fluid phase behaviors, including different scenarios for supercooled water’s highly-debated liquid-liquid transition~\cite{Gallo_Water_2016,Caupin_Thermodynamics_2019,Palmer_Metastable_2014,Buldyrev_Hamiltonian_2024}, and a fluid-fluid transition in hydrogen at extremely high pressures~\cite{Fried_Hydrogen_2022}. In most systems, where fluid polyamorphism has been attributed to the interconversion between two molecular or supramolecular states, the equilibrium constant of interconversion is a function of temperature and pressure, thus causing the fraction of interconversion to exhibit a single $T$,$P$-dependent solution in the one-phase (``disordered'') region and two $T$,$P$-dependent solutions in the two-phase (``ordered'') region~\cite{Anisimov_Polyamorphism_2018}.

However, there are examples of interconversion of species, in which the equilibrium interconversion constant does not depend on temperature and pressure, making the fraction of interconversion always 50:50 in the disordered phase. The typical example of such systems is the Ising model of a ferromagnet where the fraction of spins with the same orientation is always 50\% in the paramagnetic phase in zero field. Another example is a fluid mixture of two optical isomers, in which the chiral molecules are allowed to freely (without or with very little resistance) interconvert between two alternative configurations with equal energies, entropies, and molecular volumes~\cite{Latinwo_MolecModel_2016,Uralcan_Interconversion_2020,Petsev_Effect_2021,Wang_Chiral_2022}. If these two isomers have a greater affinity for alike molecules, the fluid, in addition to the liquid-gas transition, may exhibit a second-order transition to one of the two alternative configurations. This would be an example of ``degenerate'' single-component fluid polyamorphism, caused by the degenerate interconversion between species. 

Unrestricted molecular interconversion makes a binary mixture thermodynamically equivalent to a single-component system since the composition will no longer be an independent variable, instead becoming a function of temperature and pressure~\cite{Anisimov_Polyamorphism_2018}. If interconversion is degenerate, the alternative states become equivalent, resulting in a profound change in the formation of a new phase. Similar to magnetization in the Ising ferromagnet, the fraction of interconversion is a nonconserved property. Therefore, unlike the separation of the liquid and gas phases, the coexistence of fluid phases with the same density but with alternative molecular configurations is not required. In such systems, a new phase is growing at the expense of the alternative phase without forming a stable interface. This phenomenon, known as ``phase amplification,'' is a characteristic signature of degenerate fast interconversion (when the rate of interconversion is greater than mutual diffusion)~\cite{Uralcan_Interconversion_2020,Shum_Phase_2021,Longo_MFT_2022}. In a degenerate polyamorphic fluid, while the first-order liquid-gas transition must be accompanied by phase coexistence, the second-order liquid-liquid transition would result in the formation of a single liquid phase, resembling the Curie transition in ferromagnets or the lambda transition in superfluid helium~\cite{LL_Stat_Phys,Anisimov_Mesoscopic_2024}.

To understand the consequences of interconversion's degeneration for fluid polyamorphism, we have investigated a compressible binary lattice with unrestricted interconversion  of species, referred to as the ``blinking-checkers (BC) model'', which generally demonstrates the existence of multiple liquid-gas and liquid-liquid transitions~\cite{Caupin_Polyamorphism_2021,Longo_Interfacial_2023,Buldyrev_BCM_2024}. The degenerate version of the model was studied analytically in the meanfield approximation and computationally through 3D Monte Carlo simulations. 


The manuscript is organized as follows. First, in Section~\ref{Sec_Concepts}, non-interconverting binary mixtures that exhibit tricritical points are reviewed, and the concept of degenerate fluid polyamorphism is introduced. Then, in Section~\ref{Sec_Methods}, the BC model and the Monte Carlo (MC) simulation methods are described. The results of the meanfield calculations and Monte Carlo simulations are presented in Section~\ref{Sec_Results}, while in Section~\ref{Sec_Discussion}, the nature of the multicritical behavior observed in the BC model is discussed. Lastly, in Section~\ref{Sec_Conclusion}, concluding remarks and perspective for future studies are presented. Also, the manuscript contains an appendix describing an application of the Landau expansion to the phase transitions in the BC model, showing the emergence of tricriticality in the meanfield approximation.  

\section{Concepts}~\label{Sec_Concepts}
This section is separated into two subsections. First, a review of previous studies on symmetric binary mixtures without the possibility of interconversion is presented. Then, the concepts of symmetrical interconversion and degenerate fluid polyamorphism are introduced. 

\subsection{Multicriticality in Symmetric Binary Systems}~\label{Sec_Predictions}
Investigations of tricriticalities in real fluids and lattice models of fluids have a rich history, staring with the seminal work of Landau~\cite{Landau_Theory_1937}. Three (and more)-component fluid mixtures can exhibit so-called ``unsymmetrical'' tricritical points, in which three coexisting fluid phases become identical~\cite{Rowlinson_Liquids_1982,Furman_Global_1977}. A typical example of unsymmetrical tricriticality is the phase behavior in a ``quasibinary'' system, containing methane and a mixture of two close hexane isomers (2,2 dimethylbutan+2,3 dimethylbutane)~\cite{Knobler_Tricritical_1984}. In a mixture of methane and n-hexane, the liquid-gas critical line is interrupted by an upper critical end point (UCEP) and at a lower critical end point (LCEP). The UCEP and LCEP are connected by a line of three phase coexistence. Upon changing the pressure, the UCEP and LCEP merge into a single point, which is tricritical. Theoretically, unsymmetrical critical points may emerge due to coupling between asymmetric conserved scalar order parameters, one associated with density and the other with composition~\cite{Anisimov_Coupled_1981,Knobler_Tricritical_1984}.

In contrast, ``symmetrical'' tricritical points are observed in some binary systems, such as a ferromagnet with a nonmagnetic impurity or the superfluid $^3$He-$^4$He mixture~\cite{Vollhardt_He_1990,Schmitt_He_2015}. Symmetrical tricritical points emerge due to a coupling of a symmetric, vector-like nonconserved order parameter and a scalar conserved order parameter, associated with concentration or density. At symmetrical tricritical points, a line of second-order transitions between disordered and ordered phases becomes a line of first-order transitions, accompanied by coexistence of two phases, ordered and disordered. Investigations of some binary models, exhibiting a special symmetry with respect to the sign of the order parameter, demonstrate the possibility of symmetrical tricriticality~\cite{Hoye_SpinSystem_1976,Furman_Global_1977,Wilding_Tricritical_1995,Kofinger_Symmetric_2006}. In particular, a line of second-order phase transitions ($\lambda$ line) has been observed in the sponge phase of a fluid containing a solvent and a surfactant due to the symmetry of the system~\cite{Roux_Sponge_1992,Anisimov_Mesoscopic_2024}. 

There are several examples in the literature of binary non-interconverting systems with varying degrees of deviation from ``ideality'' (ideal-solution behavior). First, a compressible binary lattice without interconversion of species, which was thoroughly investigated by Furman et al.~\cite{Furman_Global_1977}, forms the basis of the BC model. Furman et al. demonstrate that for a certain range of interaction parameters, the non-interconverting symmetric binary mixture exhibits multicritical behavior that remarkably resembles ``symmetrical'' tricriticality. A detailed comparison between the predictions of Furman et al. and our results for the degenerate BC model is discussed in the Section~\ref{Sec_Discussion_Comparison}.

Secondly, K\"ofinger et al. investigated an off-lattice binary mixture model in meanfield and through grand canonical MC simulations~\cite{Kofinger_Symmetric_2006}. The system is similar to the degenerate BC model, except for the absence of interconversion and the implementation an off-lattice, hard-core Yukawa potential. This potential is longer-ranged than our nearest-neighbor approximation (described in Section~\ref{Sec_Methods}). In their work, the authors describe four ``topologies (or archetypes)''  of critical phase behavior: (I) A line of fluid-fluid critical points (referred to as the lambda line), in addition to the ordinary liquid-gas critical point (LGCP), intersects the liquid branch of coexistence and a critical end point,  - referred to as ``$\lambda$-end point,'' is formed; (II) two critical points are observed, one LGCP and one tricritical point (TCP); (III) only a TCP is observed; and (IV) the lambda line intersects the vapor branch of the liquid-gas coexistence and a critical point is formed (as first observed in Ref.~\cite{Scholl_TypeIV_2005}). As we will present in Section~\ref{Sec_Results}, the same four archetypes of criticality are observed in the BC model as obtained from both meanfield calculations and MC simulations.


To the best of our knowledge, there are two examples of systems that exhibit interconversion, like the BC model, and have been considered in the literature.  First is the spin fluid model, which is a compressible liquid whose particles may be in a spin up or spin down state~\cite{Hoye_SpinSystem_1976,Wilding_Tricritical_1995,Wilding_Symmetrical_1998}. In this model, the particles interact through a combination of a hard repulsion and a square-well interaction that is attractive for like-spins and opposite for unlike spins. Wilding and Nielaba~\cite{Wilding_Tricritical_1995,Wilding_Symmetrical_1998} studied this system in 2D, finding a tricritical point. Second, a ternary mixture of two optically active enantiomers containing an optically inactive substance could be considered as ``quasi-binary''~\cite{Knobler_Tricritical_1984}. It was expected that such a system could exhibit symmetrical tricriticality, like a $^3$He-$^4$He mixture. However, if the third, optically inert, component could be removed, this mixture could be viewed as a ``quasi-single-component'' system. The chiral model, investigated in Refs~\cite{Latinwo_MolecModel_2016,Uralcan_Interconversion_2020,Petsev_Effect_2021,Wang_Chiral_2022}, exhibits both liquid-gas and liquid-liquid transitions, and when interconversion of enantiomers is allowed, it behaves as a polyamorphic single-component fluid.  As shown by Wang et al.~\cite{Wang_Chiral_2022}, interconversion prevents the formation of an interface between the liquid phases in this model. The authors referred to this phenomenon as ``degenerate phase coexistence''.

\subsection{Degenerate Polyamorphism Induced by Symmetrical Interconversion}
Fluid polyamorphism ultimately originates from the complex interactions between molecules or supramolecular structures. However, this fluid behavior could be phenomenologically modeled through reversible interconversion of two alternative molecular or supramolecular states~\cite{Anisimov_Polyamorphism_2018}. The application of the ``two-state'' approach has proven to be effective in quantitatively describing liquid polyamorphism in water~\cite{Palmer_Metastable_2014,Gallo_Water_2016} and water-like microscopic models~\cite{Sastry_SingularityFree_1996,Hruby_Twostructure_2004,Ciach_Simple_2008,Stokely_Hydrogen_2010,Cerdeirina_Water_2019} as well as the fluid-fluid transition in hydrogen at extreme pressures~\cite{Fried_Hydrogen_2022}.

Interconversion of two alternative states of a molecule is equivalent to ``chemical-reaction'' equilibrium ($\ce{A <=> B}$) between two ``species,'' A and B. They can be two different structures of the same molecule (isomers), free and associated molecules, or two alternative supramolecular structures, such as different arrangements of a hydrogen-bond network. Let a variable $x$, known as the equilibrium ``reaction coordinate''~\cite{Prigogine_Chemical_1954,Anisimov_Mesoscopic_2024}, be the fraction of state B in this chemical reaction. A thermodynamic description of $\ce{A <=> B}$ interconversion can be formulated in terms of the Landau theory of phase transitions~\cite{Haar_Landau_1965,LL_Stat_Phys}, in which the order parameter, $\psi$, is associated with the equilibrium fraction of B~\cite{Longo_MFT_2022}, as $\psi = 2x-1$. The Gibbs energy per molecule is given by
\begin{equation}
    G(T,P,\psi) =G_0(T,P)+\Phi(T,P,\psi) -h\psi
\end{equation}
where $G_0(T,P) = (G_\text{A}+G_\text{B})/2$ in which $G_\text{A}$ and $G_\text{B}$ are the Gibbs energy of pure species, $\Phi(T,P,\psi)$ is the part of the free energy relevant to the phase transition, and $h = -(G_\text{A} - G_\text{B})/2$ is the ordering field conjugate to the order parameter. The equilibrium value of the order parameter is obtained by minimization of the Gibbs energy as
\begin{equation}\label{Eq_ReactionEqui}
    \pdv{G}{\psi}\bigg|_{T,P}=0
\end{equation}
resulting in $h = \partial\Phi/\partial\psi|_{T,P}$ and $\Phi$ being only a function of temperature and pressure, $\Phi = \Phi\{T,P,\psi(T,P)\}$. Since the ordering field depends only on $T$ and $P$, the equilibrium interconversion fraction is also defined by $T$ and $P$ only. 

The condition of Eq.~(\ref{Eq_ReactionEqui}) is equivalent to the chemical-reaction equilibrium condition $\partial G/\partial x|_{T,P} = 0$, while the ordering field is associated with the reaction equilibrium constant, $K$, as $h = (1/2) k_\text{B} T \ln K(T,P)$, where $k_\text{B}$ is Boltzmann's constant. The form of the function $\Phi\{T,P,\psi(T,P)\}$ depends on the interconversion conditions and interactions between the interconverting species. In the vicinity of a second-order transition, this function can be represented by a Landau expansion in powers of the order parameter in the meanfield approximation. In zero ordering field, $\ln K(T,P) = 0$, the order parameter is zero ($x=1/2$) for the disordered phase, while representing two solutions in the ordered phase, becoming a function of temperature and pressure.

\begin{figure}[t!]
    \centering
    \includegraphics[width=0.7\linewidth]{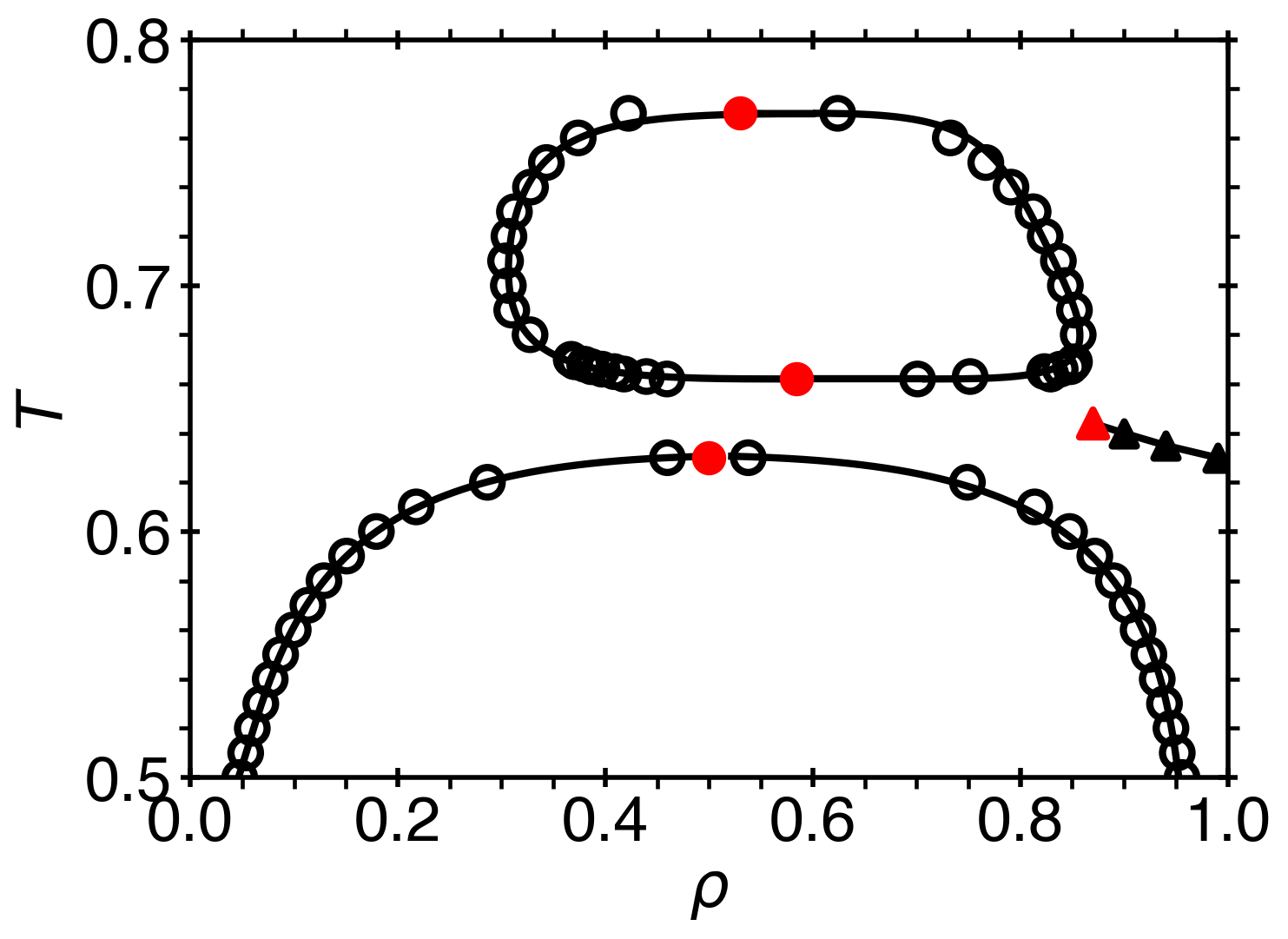}
    \caption{The $\rho$-$T$ phase diagram of a non-degenerate BC model obtained from MC simulations with specific interaction and interconversion parameters ($e=3$, $s =4$, $\omega_{11}=1.84$, $\omega_{22}=2.3$, and $\omega_{12}=1.1$, see Methods, Sec.~\ref{Sec_Methods}, for the definition of these parameters) exhibiting four critical points: three liquid-gas (circles) and one liquid-liquid (triangle). The solid lines were obtained from the complete scaling theory for asymmetric fluids~\cite{Wang_Asymmetry_2006,Wang_Asymmetry_2007,Anisimov_Mesoscopic_2024}.}
    \label{Fig_MultipleCriticalPoints}
\end{figure}

Since the fraction of interconversion is no longer an independent variable, a binary mixture with interconversion of species follows the Gibbs phase rule for a single-component substance. The interconversion of species selects a certain path, $x=x(T,P)$, through the surfaces of phase coexistence, thus crossing liquid-gas and liquid-liquid loci of the mixture at certain points. Depending on the curvature of the critical loci, the path of interconversion may cross these loci at several points exhibiting multiple individual liquid-gas and liquid-liquid critical points (see Fig.~\ref{Fig_MultipleCriticalPoints})~\cite{Longo_Interfacial_2023}. This is the origin of fluid polyamorphism.

In most polyamorphic systems, the interconversion equilibrium depends on temperature and pressure, such that the reaction equilibrium constant could be expressed as~\cite{Anisimov_Polyamorphism_2018} 
\begin{equation}\label{Eq_EquilibriumConstantGeneral}
    -k_\text{B}T\ln K(T,P) = e - Ts + P\upsilon
\end{equation}
where $e$, $s$, and $\upsilon$ are the energy, entropy, and volume change of the reaction, generally each being a function of temperature and pressure. 

There are, however, important examples of interconversion of species where $\ln K(T,P)$ is always zero, such interconversion is referred to herein as ``degenerate''. The simplest example of a system with degenerate interconversion is the Ising model for which $e$, $s$, and $\upsilon$ are zero. In zero magnetic field, the spontaneous order parameter (the magnetization, represented by the fraction of spins with the same orientation) is always $1/2$ in the paramagnetic phase, becoming a function of temperature in the ferromagnetic phase only. 

The most spectacular feature of degenerate interconversion is the phenomenon of phase amplification, when the formation of a new (ordered) phase occurs at the expense of the alternative phase. Two alternative ferromagnetic phases are different only by the sign of the spontaneous order parameter: positive or negative magnetization. This sign is randomly selected by nature (symmetry breaking) with equal probability of forming both phases. The Curie principle, stating that the symmetries of the causes are to be found in the effects, is thus obeyed, but only statistically. If the magnetization could be a conserved property, phase coexistence with the formation of the interface would be inevitable. However, because of the nonconserved nature of the Ising order parameter, the formation of a thermodynamically costly interface can be avoided. The study of a hybrid Ising model~\cite{Shum_Phase_2021}, allowing mutual diffusion of the alternative spins, demonstrated that when the interconversion of spins is significantly slower than the diffusion rate, the probability of phase amplification decreases. Without interconversion, the hybrid Ising model becomes an example of a binary alloy with coexistence of phases consisting of oppositely oriented spins.

Of course, Ising ferromagnets are not fluids; however, fast (with little resistance) degenerate interconversion, for example, has also been reported to cause phase amplification in a non-ideal mixture of enantiomers~\cite{Latinwo_MolecModel_2016,Uralcan_Interconversion_2020,Petsev_Effect_2021}. Wang et al.~\cite{Wang_Chiral_2022} referred to this effect in their ``chiral model'' as ``degenerate two-phase coexistence.'' Therefore, the chiral model describes a degenerate polyamorphic fluid exhibiting the liquid-gas transition as well as a second-order transition to one of the ordered liquid phases. Interconversion in the chiral model is unique in that the species always have equal chemical potentials. Therefore, this type of symmetric $\ce{A <=> B}$ reaction may be considered to be equivalent  to the ``vector-like'' non-conserved order parameter in the Ising model.

\begin{figure}[t!]
    \centering
    \includegraphics[width=0.4\linewidth]{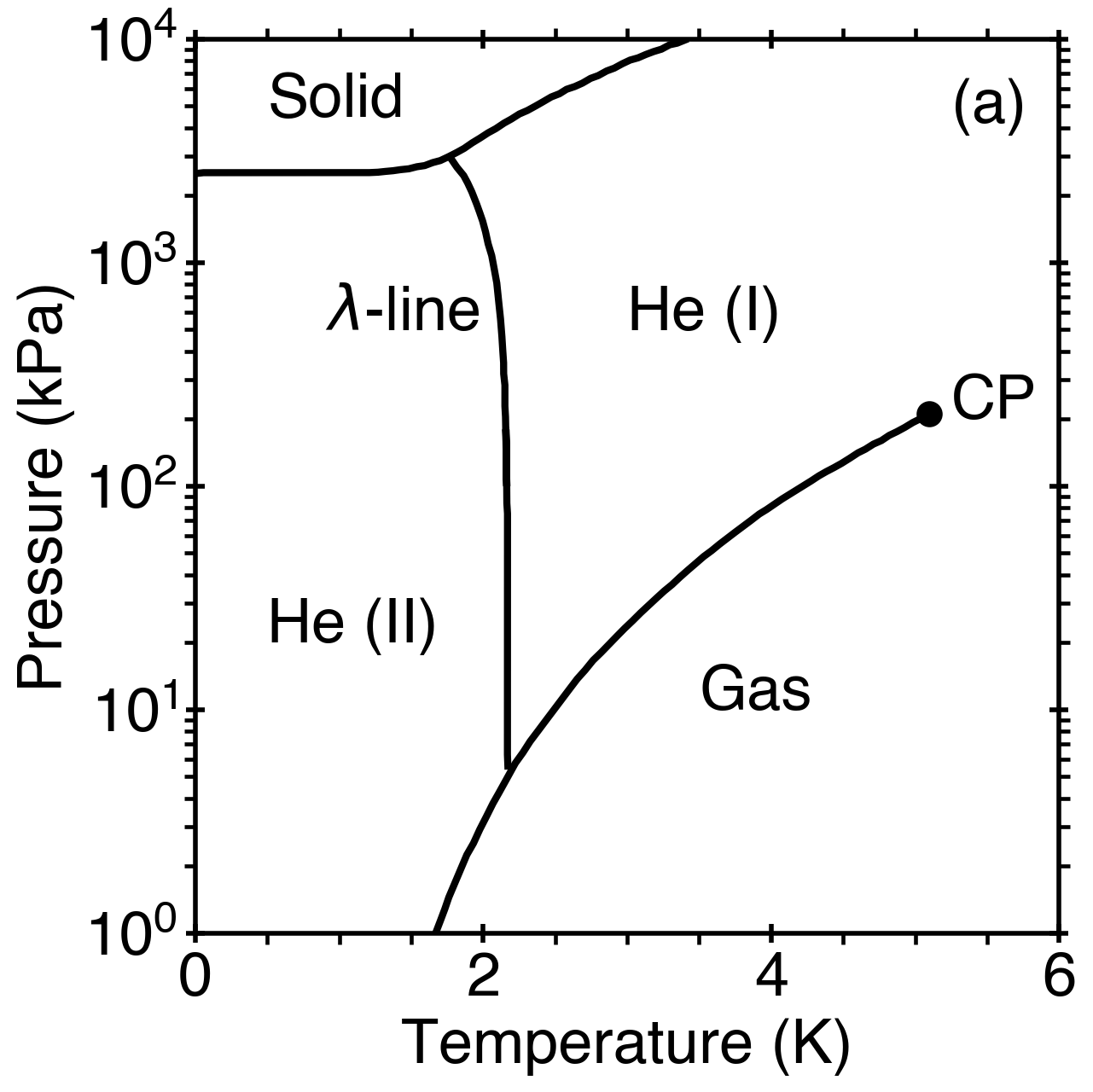}
    \includegraphics[width=0.4\linewidth]{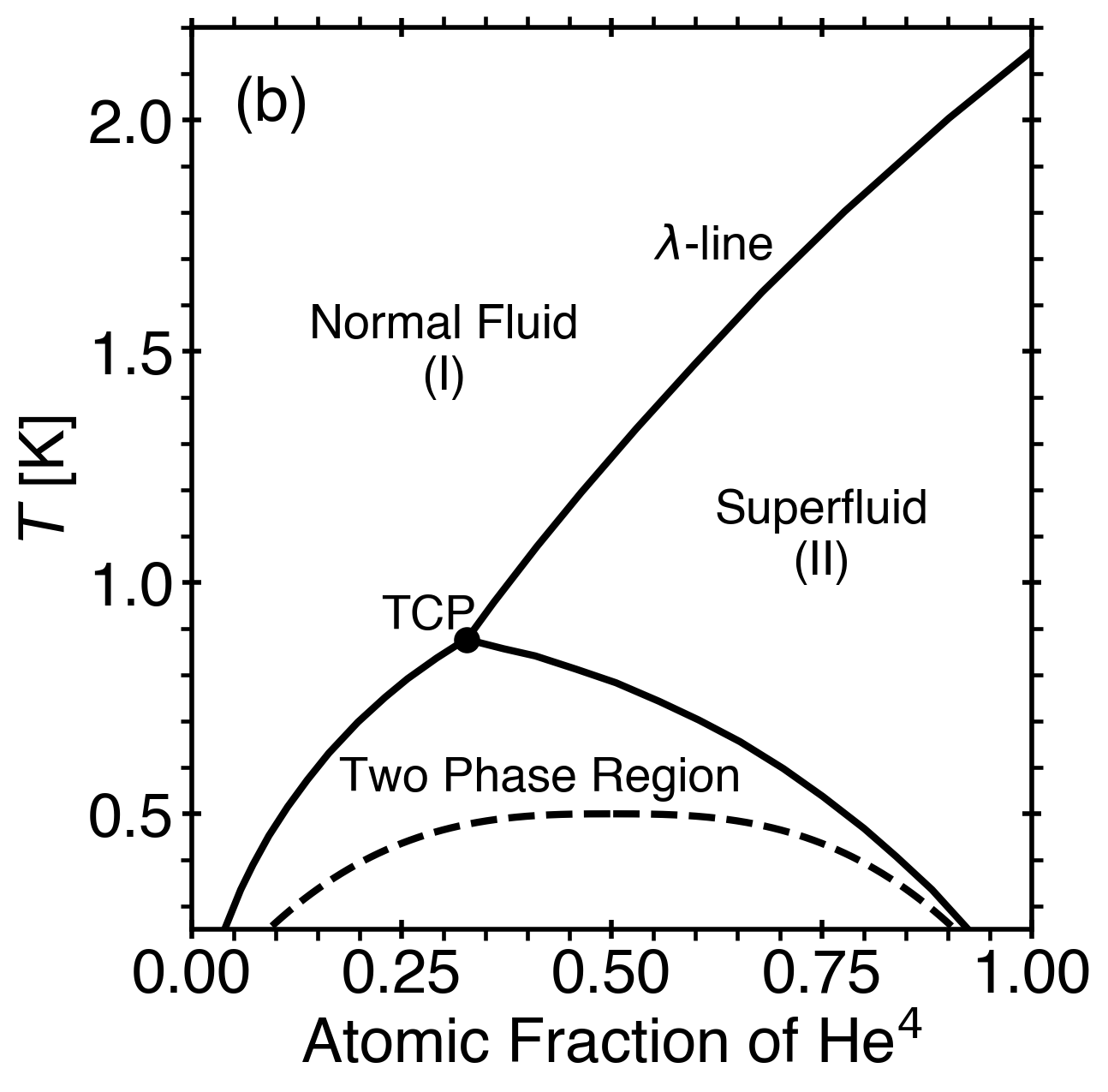}
    \caption{(a) Phase diagram of $^4$He, showing the first-order liquid-solid and liquid-gas transitions (thick lines). The liquid-gas transition is terminated at the critical point, CP). (b) Phase diagram of the liquid mixture $^3$He-$^4$He at the saturated-vapor pressure. The second-order transition to superfluidity becomes first-order at the tricritical point (TCP). Below the TCP the transition is accompanied by liquid-liquid separation. The dashed curve schematically depicts the behavior of the hypothesized liquid-liquid coexistence, uncoupled from superfluidity.}
    \label{Fig_HePhaseDiagram}
\end{figure}

A fascinating example of degenerate fluid polyamorphism is the existence of both a liquid-gas transition and a second-order transition to superfluidity in helium, which can be associated with a temperature-dependent fraction of superfluid below the lambda point~\cite{Haar_Landau_1965,Schmitt_He_2015}. The well-known $P$-$T$ phase diagram of $^4$He is schematically presented in Fig.~\ref{Fig_HePhaseDiagram}a. This figure, in addition to the liquid-gas and liquid-solid first-order phase transitions, demonstrates second-order transitions (indicated by the ``lambda-line'') in the liquid phase, which is terminated by crystallization and evaporation. 

The existence of a second-order transition line in the liquid phase is a conceptual consequence of degenerate interconversion. In non-degenerate polyamorphic systems, as interconversion is a function of $T$ and $P$, the interconversion path selects individual points on the fluid-fluid critical lines, while in the case of degenerate interconversion, a pressure-dependent line of order-disorder phase transitions emerges. Such a line of second-order transitions between two fluid states has been observed in the chiral model of interconverting enantiomers~\cite{Uralcan_Interconversion_2020,Wang_Chiral_2022}. 


As seen in Fig.~\ref{Fig_HePhaseDiagram}a, the lambda line in $^4$He is located far away from its liquid-gas critical point. Therefore, these two criticalities are essentially independent. This means that the two order parameters, one (vector-like) associated with the quantum wave function of the superfluid state and another (scalar), associated with the molecular density, are uncoupled. However, in a mixture of helium isotopes, $^4$He and $^3$He, the lambda transition at a certain concentration becomes close to the uncoupled metastable critical point of demixing between these two components (illustrated by Fig.~\ref{Fig_HePhaseDiagram}b). 

In this system, the two order parameters vector and scalar (associated with the mixture concentration) are coupled, generating a first-order transition to superfluidity. The point, where the second-order transition becomes first order, is known as a ``symmetric tricritical point''. Note that this is not the same as the point where three fluid phases become critical, which is referred to as an ``asymmetric tricritical point'' or ``fluid tricritical point''~\cite{Rowlinson_Liquids_1982,Lawrie_Tricritical_1984,Knobler_Tricritical_1984}. Near the tricritical point, the coexistence between two liquid states exhibits a characteristic angle-like shape. This feature of tricriticality is illustrated in Figure~\ref{Fig_HePhaseDiagram}b. The angle-like shape of the coexistence of the solvent and polymer solution in the limit of infinite degree of polymerization near the Theta point is one of the factors proving that the Theta point is a tricritical point~\cite{Anisimov_Mesoscopic_2024,Anisimov_PolymerTricrit_2024}.

In this work, we studied both analytically (in the meanfield approximation) and computationally (via 3D Monte Carlo simulations) a degenerate version of a minimal microscopic model for fluid polyamorphism - the BC model (described in detail in the next section). We have observed and investigated a novel phenomenon for single-component polyamorphic fluids: the emergence of symmetric tricritical points, analogous to the tricritical point in the superfluid mixture of helium isotopes.

\section{Methods}~\label{Sec_Methods}
In this section, functions and parameters of the meanfield degenerated BC model are provided, while the Monte Carlo (MC) simulation details are elaborated.  

\subsection{Blinking-Checkers (BC) Model~\label{Sec_Methods_BCM}}
The non-degenerate BC model has been investigated in Refs.~\cite{Caupin_Thermodynamics_2019,Longo_Interfacial_2023,Buldyrev_BCM_2024}, in which there are five free parameters that affect the free energy per lattice site: $e$, $s$, $\epsilon_{11}$, $\epsilon_{22}$, and $\epsilon_{12}$. The first two are the energy and entropy change of the interconversion reaction, as described in Eq.~\ref{Eq_EquilibriumConstantGeneral}, while the latter three are the interaction parameters for species of type 1 with itself, type 2 with itself, and the cross interaction between species of type 1 and type 2. In this work, we consider a  degenerate version of the BC model, in which $\ln K = 0$ (see Eq.~\ref{Eq_EquilibriumConstantGeneral}) and $\epsilon_{11}=\epsilon_{22}$. Introducing, $\omega_{11}=-Z_n\epsilon_{11}/2$, $\omega_{11}=-Z_n\epsilon_{11}/2$, and $\omega_{11}=-Z_n\epsilon_{11}/2$, where $Z_n$ is the number of nearest neighbors, the Helmholtz free energy per unit volume, $f$, is given by
\begin{equation}~\label{Eq_BCfreeEnergy}
\begin{split}
    f(T,\rho,x) =& \rho\varphi_2^\circ - \frac{\rho^2}{2}\left[\omega_{11} + \omega_{11}(2x-1)^2 + 4\omega_{12}x(1-x)\right]\\ &+ T\left[\rho x \ln{x} + \rho(1-x)\ln(1-x)\right] + T\left[\rho\ln\rho + (1-\rho)\ln(1-\rho)\right]
\end{split}
\end{equation}
in which $k_\text{B}=1$ and $\varphi_2^\circ=\varphi_2^\circ(T)$ is a function of temperature only, containing the arbitrary zero points of energy and entropy\cite{Anisimov_Polyamorphism_2018}, while $\rho$ is the number density and $x$ is the molecular fraction of species. The second through fourth terms in square brackets in Eq.~(\ref{Eq_BCfreeEnergy}) describe the contribution to the free energy from the energy of interactions, the entropy of mixing of the two species, and the entropy of mixing of the occupied and empty sites. We note that a ``symmetric'' version of the non-interconverting BC model (where $\omega_{11}=\omega_{22}$) was investigated in the supplemental material of Ref.~\cite{Caupin_Polyamorphism_2021}. 

The condition for chemical-reaction equilibrium, $\partial f/\partial x|_{T,\rho}=0$ constrains the equilibrium concentrations (molecular fractions) of species 1, $x=x_\text{e}(T,\rho)$, such that the interconverting binary mixture thermodynamically behaves as a single component fluid. For the degenerate BC model, applying this condition to Eq.~(\ref{Eq_BCfreeEnergy}) yields,
\begin{equation}\label{Eq_RhoFunctionOfX}
    \rho = \frac{T}{2(\omega_{11}-\omega_{12})(1- 2x)}\ln\left(\frac{1-x}{x}\right)
\end{equation}
which, at fixed $\rho$, gives two solutions for $x$ in the liquid phase and one solution for $x$ in the gaseous phase. 

The chemical potential is obtained from the derivative of the free energy along equilibrium as $\mu = \partial f\left[T,\rho, x(T,\rho)\right]/\partial\rho|_T$. Phase equilibrium is obtained by numerically inverting Eq.~(\ref{Eq_RhoFunctionOfX}) to get $x=x(\rho,T)$, then the conditions of equal pressure and chemical potential were used to obtain the liquid-gas coexistence. The limits of thermodynamic stability (spinodal curves) were determined from $\partial\mu/\partial\rho|_T=0$, which yields an analytic equation. The fluid-fluid critical point was obtained from the location where the spinodal reaches a smooth (``parabolic'') maximum, while the tricritical point was obtained from the location where the spinodal exhibits a cusp at its maximum. 

In the degenerate BC model, the liquid-gas critical parameters for the temperature and density are $T_\text{c} = (\omega_{11}+\omega_{12})/2$ and $\rho_\text{c}=1/2$, respectively. To make an appropriate comparison with the MC data, the temperature in both data sets were normalized such that $\hat{T} = 2T/\omega_{11}$. Lastly, the lambda line is obtained where $x=1/2$ and $\hat{T}_{\lambda} = 2\rho(1-\omega_{12}/\omega_{11})$, see more details in Ref.~\cite{Caupin_Thermodynamics_2019}. For this reason, we define a reduced species-to-species interaction parameter, $\bar{\omega}$, as 
\begin{equation}
    \bar{\omega} = 1 - \frac{\omega_{12}}{\omega_{11}}
\end{equation}
The case $\bar{\omega} = 0$ ($\omega_{12}=\omega_{11}$) corresponds to the liquid-gas transition, which has no second-order phase transition between the two liquid phases, which we refer to this as the ``uncoupled'' liquid-gas transition. Note that $T_\text{c} = 1$ when $\bar{\omega}=0$.

The pressure is determined from $P = \rho\mu-f$ and is given by
\begin{equation}~\label{Eq_MF_pressure}
    \hat{P} = -2\rho^2\left[1 -2\bar{\omega}x(1-x)\right] - \hat{T}\ln(1-\rho)
\end{equation}
where the pressure is normalized just as the temperature, such that $\hat{P}=2a^3 P/\omega_{11}$ where $a$ is the size of a lattice cell. We define $a=1$, so that $\rho$ is the number density. Along the lambda line, $x=1/2$ and $\hat{T}_\lambda=2\rho\bar{\omega}$, such that the pressure becomes a function $\hat{P}_\lambda=\hat{P}(\rho,\bar{\omega})$.

\subsection{Monte Carlo Simulations~\label{Sec_Methods_MCSimulations}}

Monte Carlo simulations are performed on a cubic lattice in an elongated box with $L_x=L_y=128$ and $L_z=256$, such that the total number of lattice sites is $n=L_xL_yL_z=2^{22}\approx 4\times 10^6$. Each lattice site, $i$, interacts with it's $Z_n=6$ nearest neighbors or $Z_n=26$ nearest and next-to-nearest neighbors (within the geometric distance smaller than two lattice constants) and may be in one of three possible states: $s_i=0$ (empty), $s_i=1$ (particle of type 1), and $s_i=2$ (particle of type 2). The number of empty sites, $n_0$, is fixed, while the number of particles of types 1 and 2 (given by $n_1$ and $n_2$, respectively) is allowed to vary through an interconversion reaction, such that $n_0+n_1+n_2=n$. The density and mole fractions are defined as $\rho=(n_1+n_2)/n$, $x_1=n_1/(n_1+n_2)$, and $x_2=n_2/(n_1+n_2)$. The potential energy of each site is computed as $u_i=\sum_j^{Z_n}\epsilon(s_{i},s_{ij})$, where $\epsilon(k,l)=2\omega_{kl}/Z_n$ is a symmetric matrix with $\omega_{0k}=\omega_{k0}=0$. Further details on the simulations can be found in Refs.~\cite{Longo_Interfacial_2023,Buldyrev_BCM_2024}. In this work, the degenerate BC model was simulated for $e=s=0$ and $\omega_{11} = \omega_{22}$.

For the system with interconversion, at each Monte Carlo step, a Kawasaki swap\cite{kawasaki_diffusion_1966} (simulating diffusion) and a Glauber flip\cite{glauber_timedependent_1963} (simulating an interconversion reaction\cite{Shum_Phase_2021}) are attempted. The simulations can also be performed with interconversion or without (only Kawasaki swaps). Simulations with interconversion are initialized with $x=1$, while simulations without interconversion are initialized with $x=1/2$. In accordance with the Metropolis criterion\cite{metropolis_basic_1963}, the new state is accepted with probability $p=\exp(-\Delta F/T)$, where $\Delta F$ is the change in the free energy after the swap or flip. More details on how interconversion occurs in this system can be found in Refs.~\cite{Shum_Phase_2021,Longo_Formation_2022,Longo_Interfacial_2023,Buldyrev_BCM_2024}. The configurations are saved in a compressed format in which each lattice site is encoded in two bits to keep the values 0, 1, and 2 for the particles of different types. Thus, each configuration file uses $n/4$ bites. The stored configurations are then analyzed to find density and mole-fraction profiles for a planar slab. 

The system is initialized such that cells at locations $0 \leq z<L_z\rho  x_1$ are occupied with particles of type 1, cells at locations $L_z\rho  x_1\leq z<L_z\rho$ are occupied with particles of type 2, and cells at locations $z\geq \rho L_z$ are empty. The system is equilibrated for $2^{13}$ time units (one time unit consists of $n$ Monte Carlo moves). After equilibration, if the system is above the critical point, a supercritical fluid state may be observed, while below the critical point, two fluid phase coexistence is found. A consequence of interconversion is that, along coexistence, the liquid phase may interconvert into a system that is primarily made up of particles of type 1 or type 2. This occurs to avoid the energy penalty of forming an interface between species that have an unfavorable interaction energy (see more details on this phenomenon, referred to as ``phase amplification'', in Refs.~\cite{Shum_Phase_2021,Longo_Formation_2022}).

\begin{figure}[t!]
    \centering
    \includegraphics[width=0.4\linewidth]{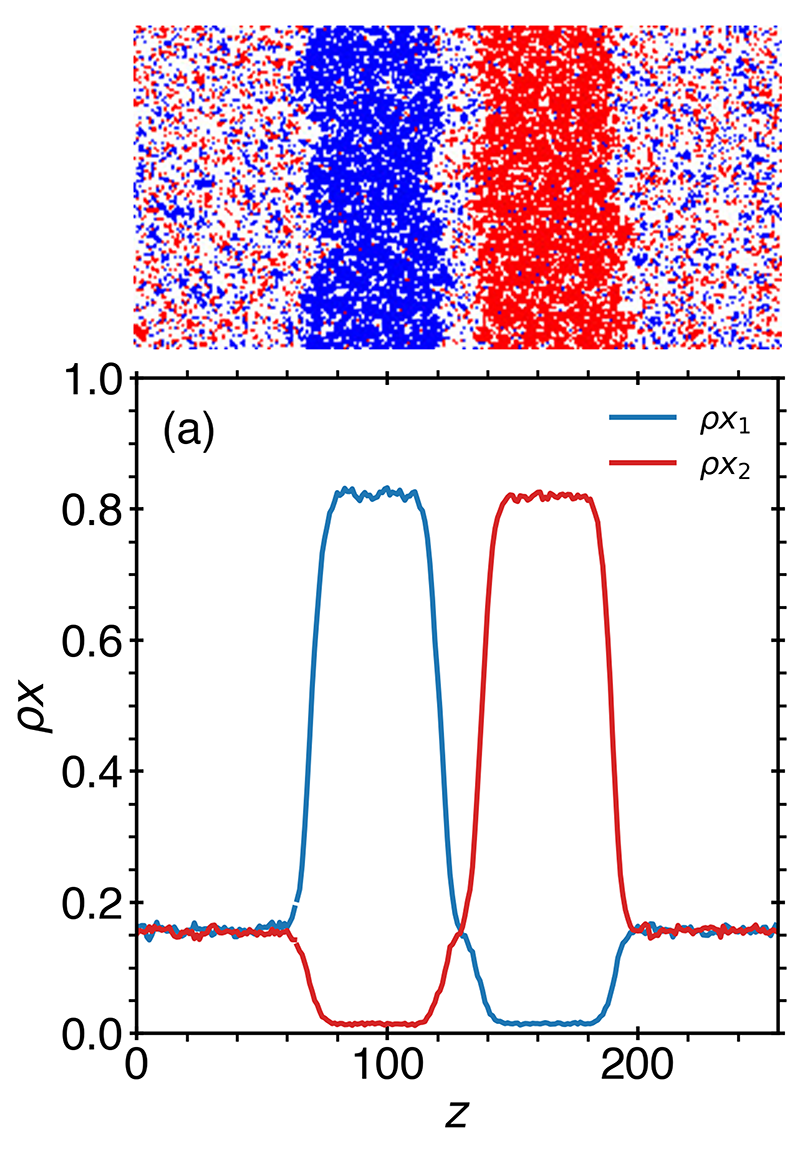}
    \includegraphics[width=0.4\linewidth]{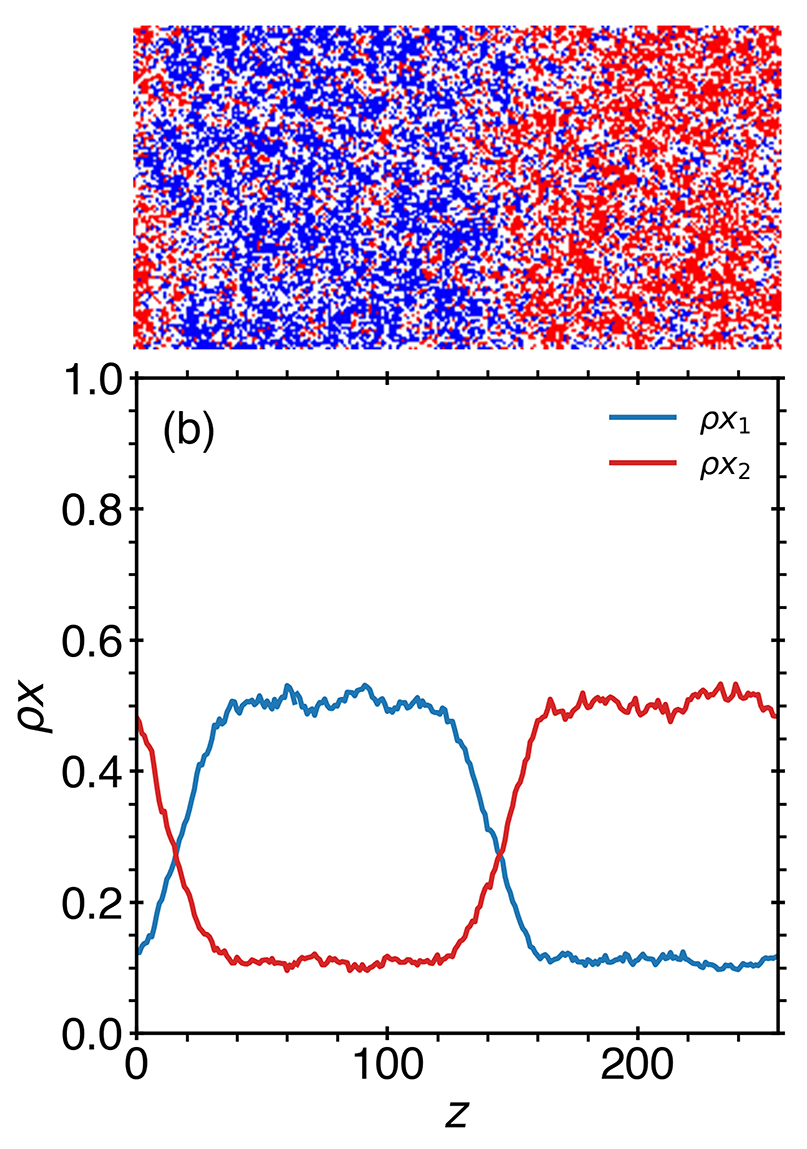}
    \caption{Partial density ($\rho x$) profiles of the BC model without the possibility for interconversion, obtained from MC simulations with $\bar{\omega}=1$ at time $t=2^{13}$ MC time steps. In (a), a state along coexistence is shown for $\hat{T}=0.72$, average $\rho = 0.52$, and average $x=0.5$, while in (b) a supercritical liquid-gas state is shown for $\hat{T}=0.85$, average $\rho =0.6$, and average $x=0.5$. Above the profiles are simulation snapshots, depicting a 2D slice along the $z$-direction, of the system where the blue particles are liquid 1, red are liquid 2, and white indicate gas.}
    \label{Fig_Snapshots_NoInt}
\end{figure}

\begin{figure}[t!]
    \centering
    \includegraphics[width=0.4\linewidth]{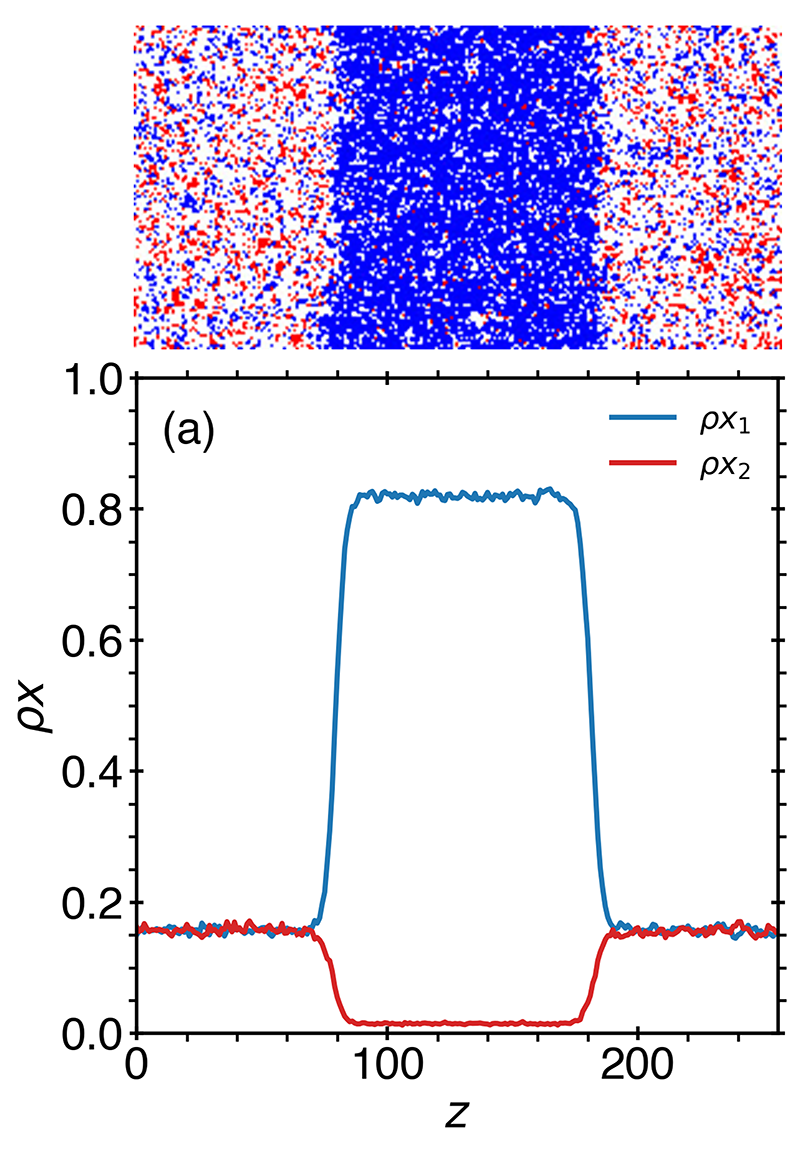}
    \includegraphics[width=0.4\linewidth]{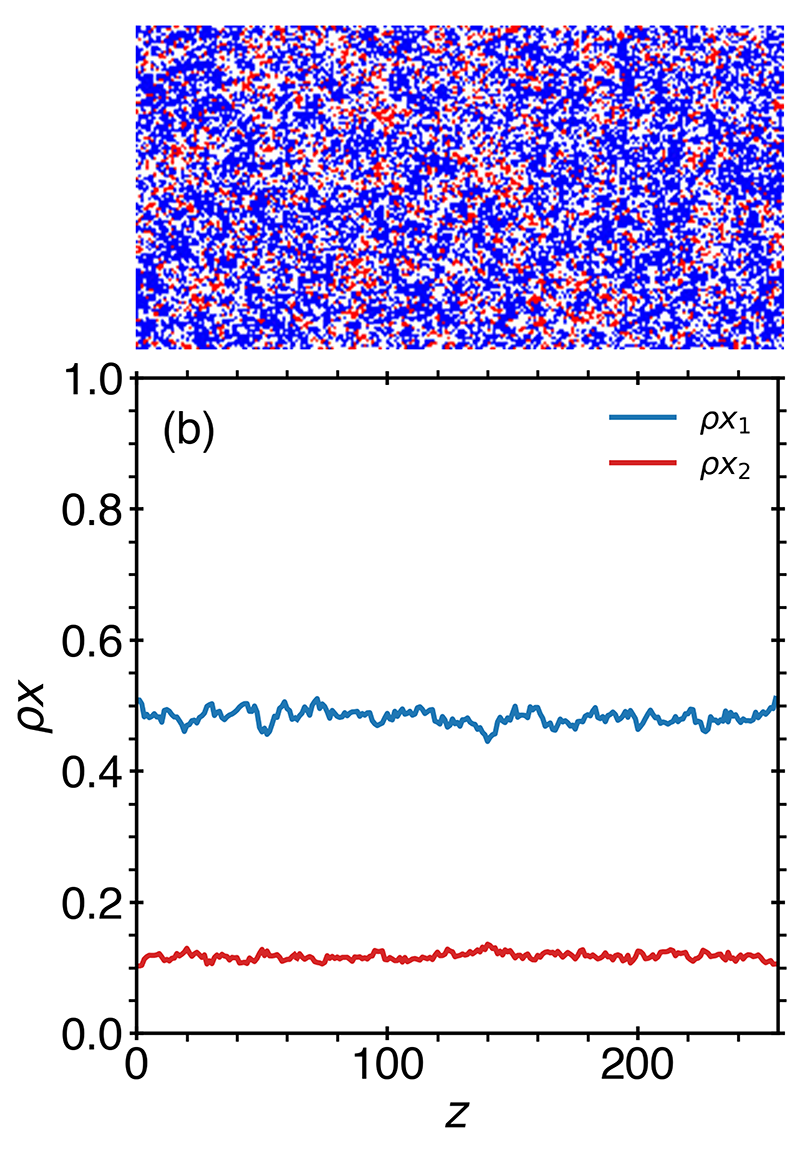}
    \caption{Partial density ($\rho x$) profiles and snapshots of the BC model with the possibility of interconversion, obtained from MC simulations for the same $\bar{\omega}$, $\rho$, $T$, and time as presented in Fig.~\ref{Fig_Snapshots_NoInt}. When interconversion is possible, one of the two liquid phases are amplified to avoid the energetically unfavorable interface, such that in equilibrium, only one of the two phases remain.}
    \label{Fig_Snapshots_Int}
\end{figure}

Figures~\ref{Fig_Snapshots_NoInt} and ~\ref{Fig_Snapshots_Int} show the equilibrium behavior (at time $t=2^{13}$) of two sample systems with and without interconversion, respectively, at $\bar{\omega}=1$. Note that in both figures panels (a) show a system below $T_\text{c}$, while panels (b) show a system above $T_\text{c}$ for the same set of parameters. The effect of phase amplification is shown upon comparison of the system with and without interconversion. Also, note that in Fig.~\ref{Fig_Snapshots_NoInt}a, the phase consisting of particles of type I (blue) and type II (red) are separated by a small gaseous phase. This is a wetting effect as the interactions between the two species are so unfavorable, the gaseous phases wets the two liquids to reduce the overall free energy. This effect was discussed in detail in Ref.~\cite{Longo_Interfacial_2023}.

In order to characterize the two phases quantitatively, the average fractions of particles $x_1(z)$, $x_2(z)$, and the density $\rho(z)$ were computed by calculating the numbers of cells with given coordinate $z$ occupied by particles of type 1 ($n_1(z)$), type 2 ($n_2(z)$), and empty cells. In Figs.~\ref{Fig_Snapshots_NoInt} and ~\ref{Fig_Snapshots_Int}, the concentration profiles for the configurations, represented by slices (a) and (b), are shown in panels (c) and (d), respectively.

The equilibrium density and fraction of particles of all phases are computed from their profile, obtained with interconversion, as a function of the horizontal coordinate $z$ and by the global minimum and maximum of $\rho(z)$, $x_1(z)$, and $x_2(z)$. These functions were averaged over a window of length $\Delta z=64$, centered around the positions of the global minimum and maximum. The equilibrium temperature-density diagrams are presented in the next Section. The error bars were determined as half the difference between the average densities in the low and high density phases, computed near the liquid-gas critical point.

\section{Results}~\label{Sec_Results}
In this Section, we present the phase behavior of the degenerate BC model in agreement with the predictions of Furman et al.~\cite{Furman_Global_1977} and K\"ofinger et al.~\cite{Kofinger_Symmetric_2006} for their non-interconverting symmetric binary mixtures. The section is organized into two parts. In Section~\ref{Sec_Results_TRho_Data}, we present temperature-density phase diagrams, split into four subsections categorized by the four archetypes of tricriticality as defined by K\"ofinger et al., while in Section~\ref{Sec_Results_PT_Data} we compare the pressure-density and pressure-temperature phase diagrams for each archetype.

\subsection{Temperature-density phase diagrams}~\label{Sec_Results_TRho_Data}
This subsection depicts meanfield results for each archetype of tricriticality. Furman et al.~\cite{Furman_Global_1977} predicted that the boundaries of these four regions for the degenerate BC model are at $\bar{\omega}=0.417$ (the onset of tricriticality), $\bar{\omega}=0.538$ (the end of type II behavior), and $\bar{\omega}=(4/3)(5-\sqrt{10})\approx2.450$ (the disappearance of tricriticality). The archetypes are presented in order of increasing $\bar{\omega}$. Monte Carlo simulations are also shown in comparison for similar values of $\bar{\omega}$. The results are organized by archetype, but as we shall demonstrate, the possibility of observing Type II or Type IV behavior in the MC simulations depends on the coordination number, $Z_n$.

\subsubsection{Type I Behavior}
\begin{figure}[b!]
    \centering
    \includegraphics[width=0.4\linewidth]{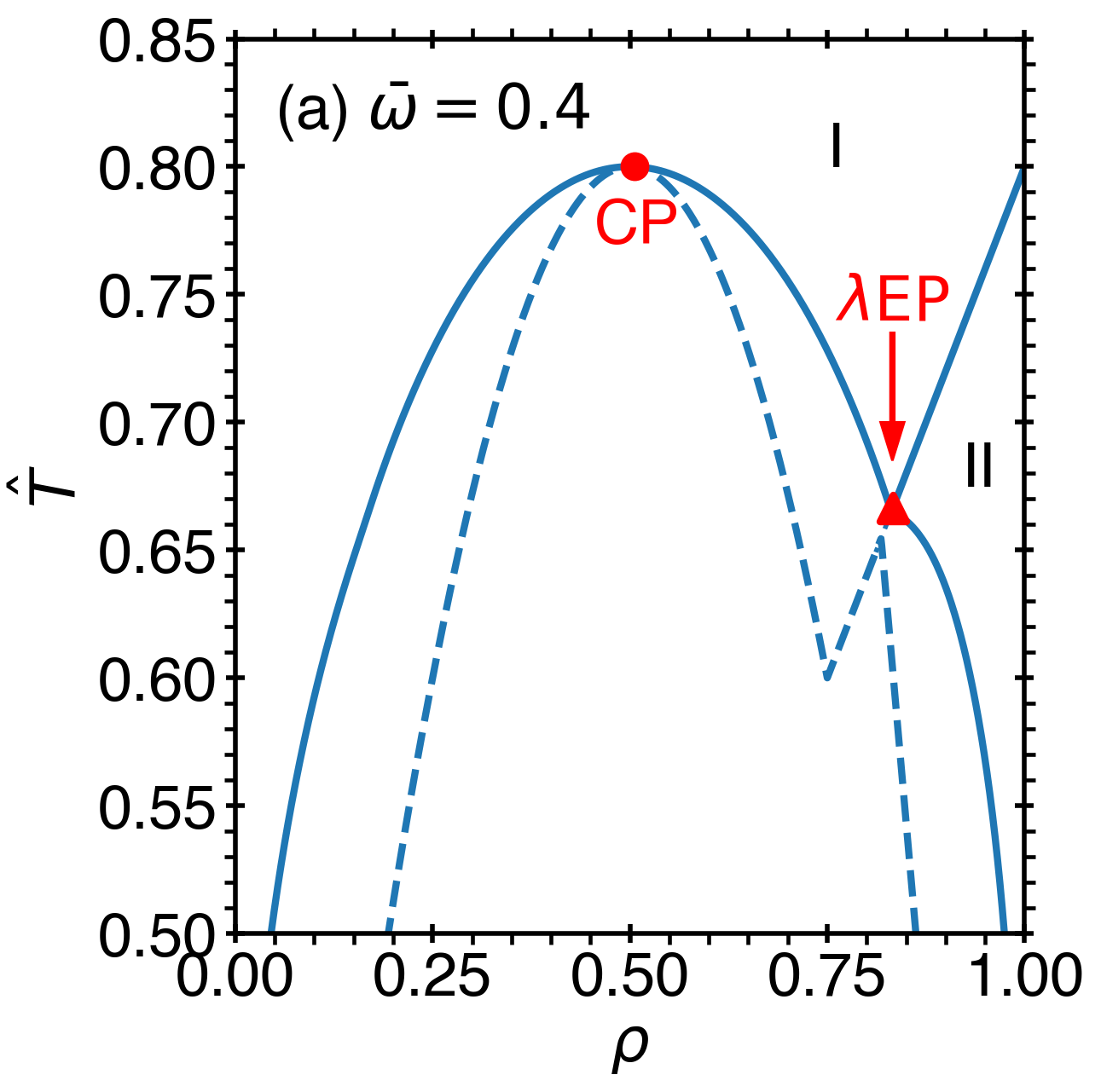}
    \includegraphics[width=0.4\linewidth]{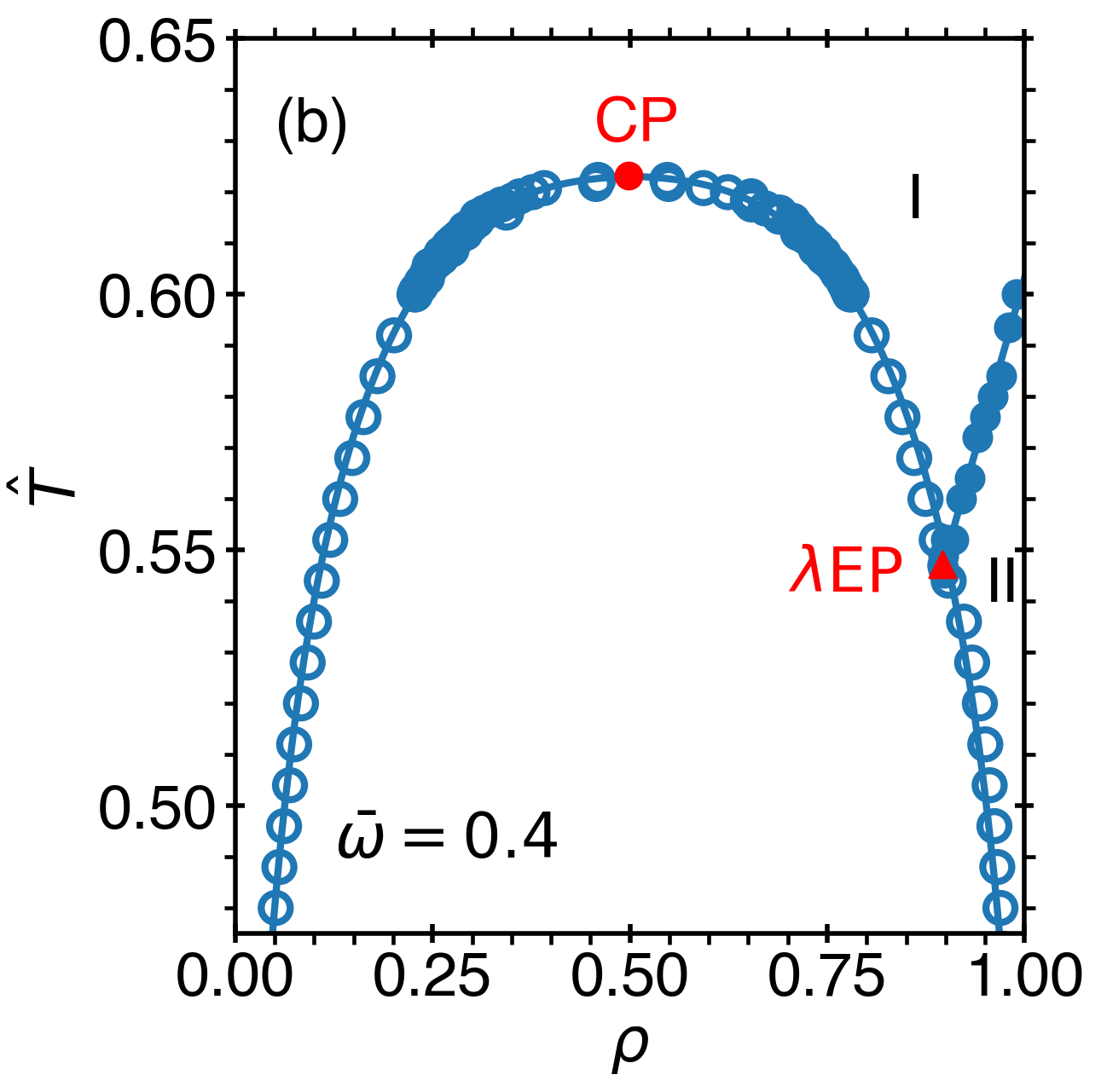}
    \caption{Type I behavior. Comparison of meanfield and Monte Carlo (MC) phase diagrams for $\bar{\omega}=0.4$ and $Z_n=6$ in the degenerate BC model showing type I behavior. In (a) the solid curve corresponds to liquid-gas coexistence, while the dashed curve represent the stability limit (spinodal). In (b), for the MC simulation data, the open circles correspond to the liquid-gas coexistence, the closed circles illustrate the lambda line, while the solid curve represents guidelines for the coexistence. The area marked as ``I'' corresponds to the ``disordered'' fluid, while the area marked as ``II'' corresponds to the ``ordered'' fluid where interconversion depends on temperature and density. The line that separates these two regions is the ``lambda'' line. The lambda line crosses the coexistence at a lambda-end point, $\lambda\text{EP}$ (triangle).}
    \label{Fig_TypeIbehavior}
\end{figure}

As discussed in Section~\ref{Sec_Predictions}, for $\bar{\omega}<0.417$, type I behavior is characterized by systems that exhibit a lambda line that intersects with the liquid branch of coexistence below the liquid-gas critical point. This is observed in the degenerate BC model for $\bar{\omega}=0.4$ and is presented in Fig.~\ref{Fig_TypeIbehavior} as obtained by both meanfield calculations and MC simulations. 

Analogously to liquid-liquid critical end points in binary fluid mixtures~\cite{Rowlinson_Liquids_1982}, we refer to the interception of the lambda line and the liquid-gas coexistence line as a ``$\lambda$-end'' point. Note that, for $\bar{\omega}=0.4$, one also observes an inflection in the liquid branch of the coexistence and a cusp in the liquid-gas spinodal in the meanfield calculations (Fig.~\ref{Fig_TypeIbehavior}a). This cusp is a result of the intersection of the lambda line with the liquid-gas coexistence, and thus, it may be interpreted as a metastable or ``virtual'' tricritical point. 

Monte Carlo simulations are presented in Fig.~\ref{Fig_TypeIbehavior}b, in which the open circles represent the simulation data for liquid-gas coexistence. The solid curve illustrates the prediction for the crossover from meanfield behavior far away from the critical point to the asymptotic scaling power law: $\varphi = 2\rho-1 \propto \Delta \hat{T}^\beta$ where $\Delta\hat{T} = |T-T_\text{c}|/T_\text{c}$ and $\beta=0.326$ in the 3D Ising universality class of criticality~\cite{panagiotopoulos_direct_1987,Luijten_Medium_1996,Luijten_Nature_1998}. 

\subsubsection{Type II Behavior}

\begin{figure}[b!]
    \centering
    \includegraphics[width=0.4\linewidth]{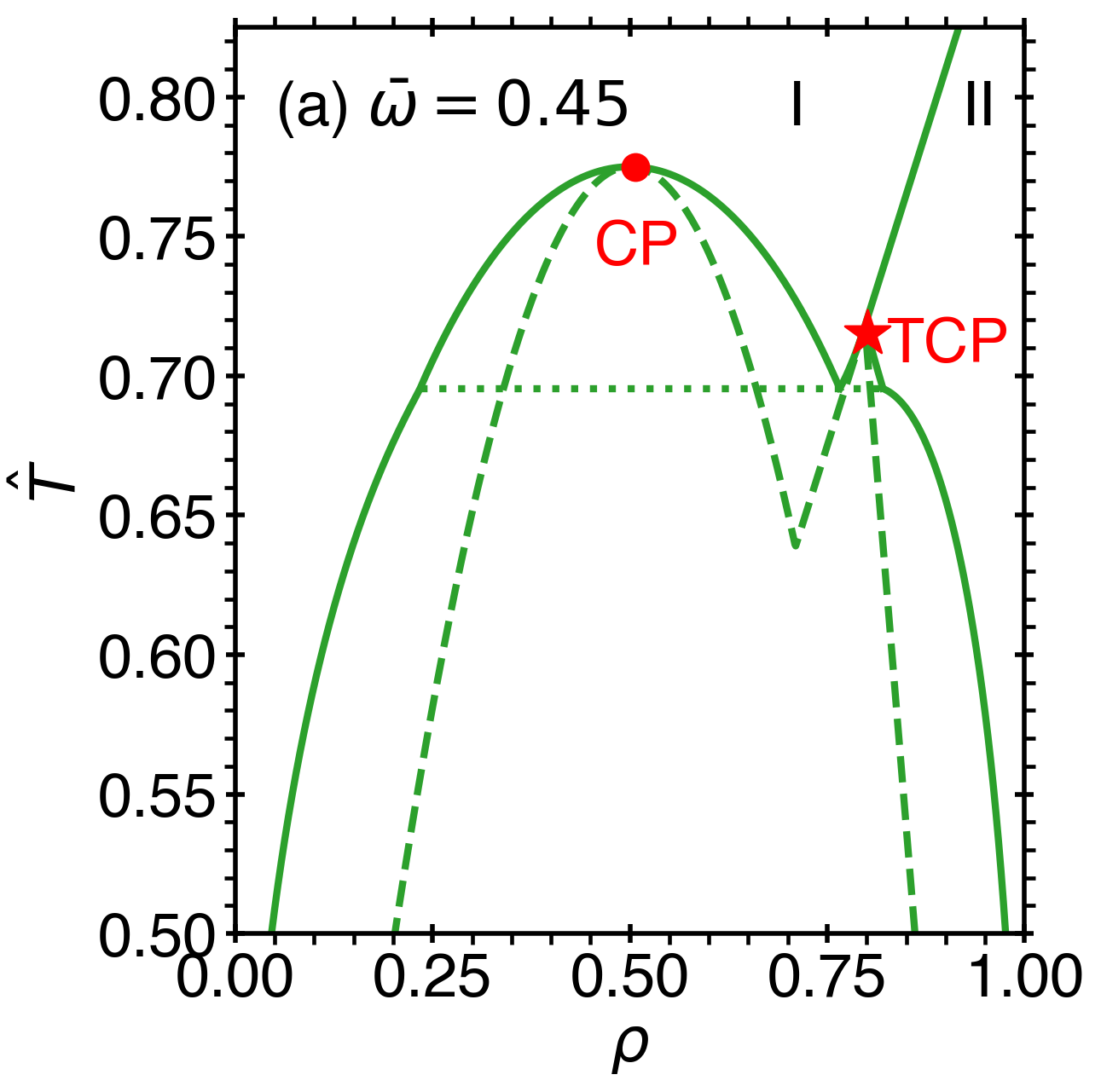}
    \includegraphics[width=0.4\linewidth]{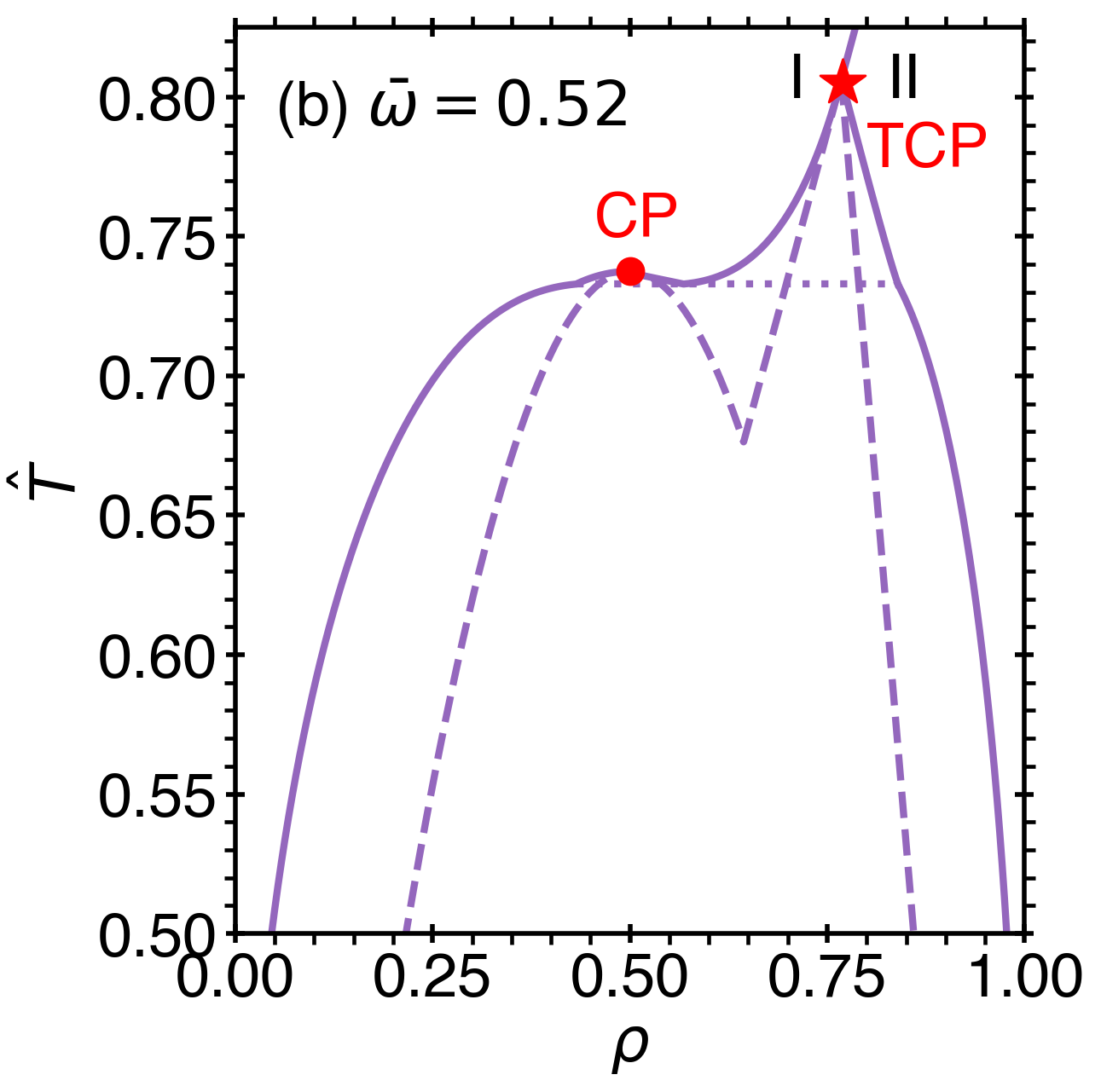}
    \includegraphics[width=0.4\linewidth]{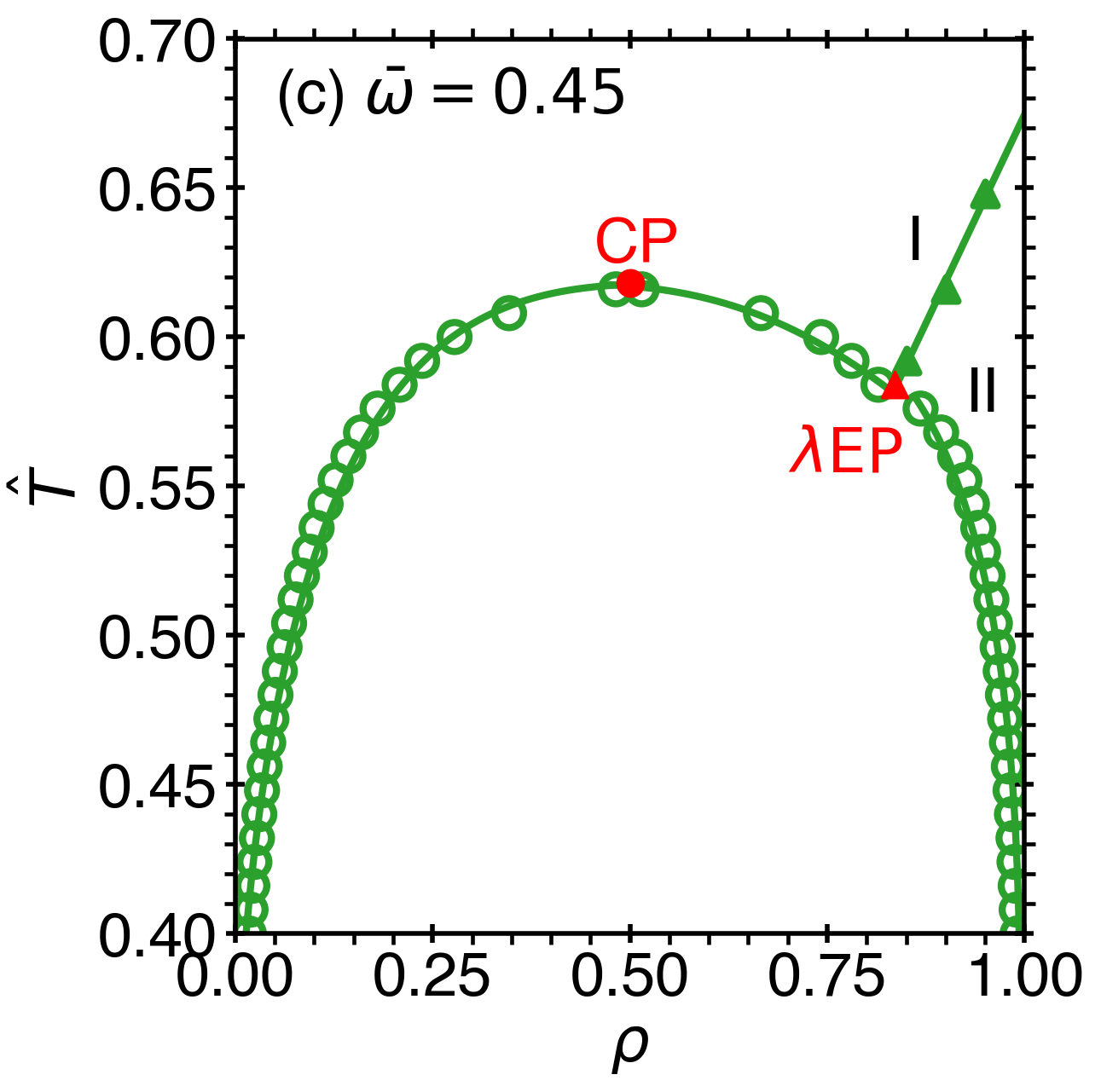}   
    \includegraphics[width=0.4\linewidth]{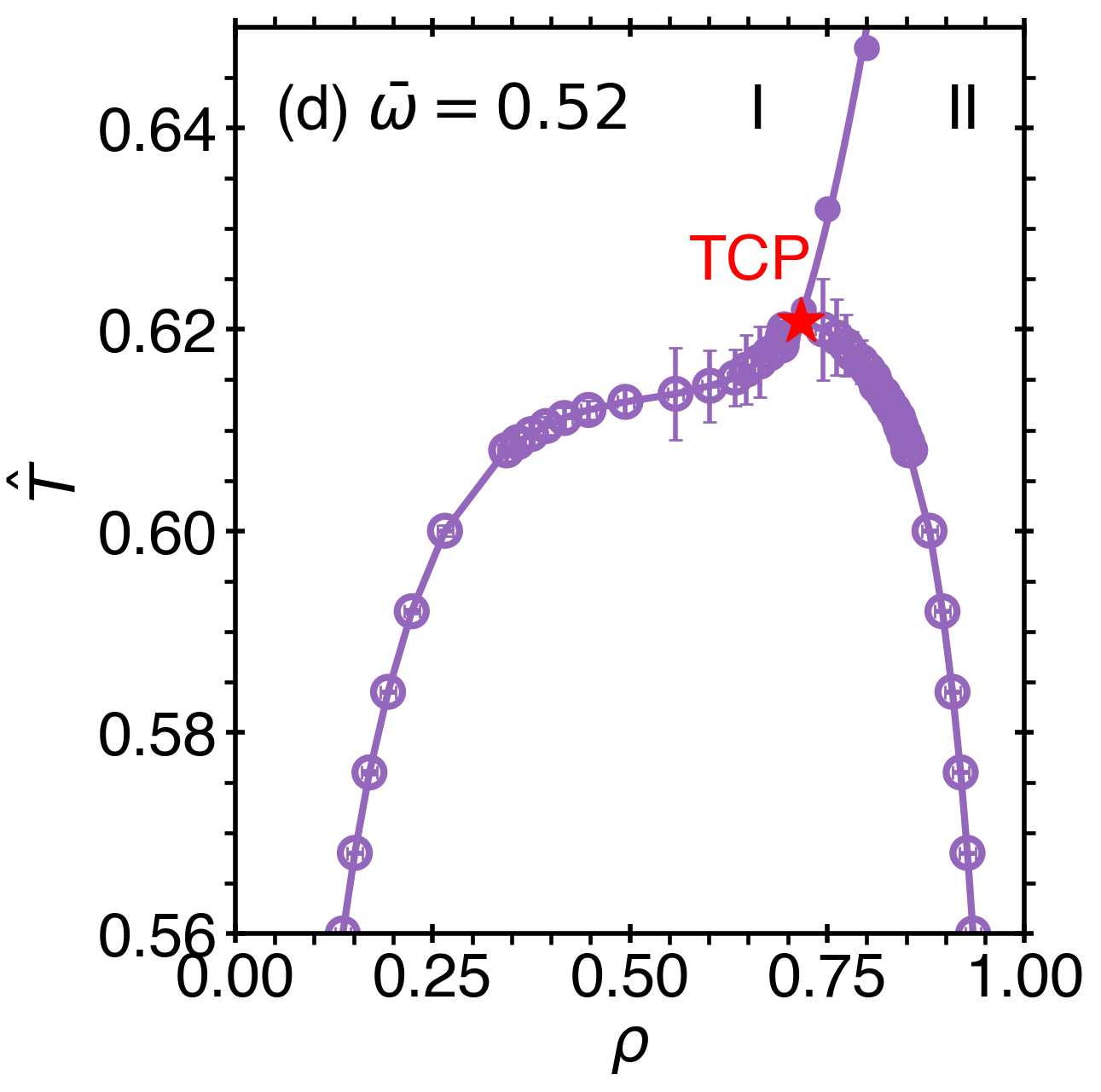}   
    \caption{Phase II behavior. Comparison of meanfield and MC (with $Z_n=6$) temperature-density phase diagrams for two values of $\bar{\omega}=0.45$ (a,c, green) and $\bar{\omega}=0.52$ (b,d blue) in the degenerate BC model showing type II behavior. In (a,b) curves correspond to liquid-gas coexistence (solid), the stability limit (dashed), the triple temperature (dotted). MC simulation data is presented in (c,d), where the open circles correspond to the liquid-gas coexistence, the closed circles illustrate the lambda line, while the solid curve represents guidelines for the coexistence. The area marked as ``I'' corresponds to the ``disordered'' fluid, while the area marked as ``II'' corresponds to the ``ordered'' fluid where interconversion depends on temperature and density.}
    \label{Fig_App_wBar_0p5}
\end{figure}

Type II behavior is observed in the interval $0.417\le \bar{\omega} \le 0.538$ and is characterized by systems that exhibit two critical points: a liquid-gas critical point (LGCP) and a tricritical point (TCP). The TCP also indicates a liquid-liquid critical point between a high-density ordered liquid and a low-density disordered liquid, each composed of particles of type 1 or 2, at which the mole fractions of both particles are equal. Figure~\ref{Fig_App_wBar_0p5} illustrates a comparison of meanfield and MC simulations in the degenerate BC model that confirm type II behavior in this region. The meanfield calculations, presented in Fig.~\ref{Fig_App_wBar_0p5}(a,b) for $\bar{\omega}=0.45$ and $\bar{\omega}= 0.52$, demonstrate the two critical points predicted for this archetype. The system also exhibits a vapor-disordered liquid-ordered liquid triple point at a temperature less than CP and TCP. Note that in the degenerate BC model, the disordered liquid and ordered liquid are comprised of either liquid-1 or liquid-2 as, due to interconversion, phase amplification causes one of these two phases to grow at the expense of the other to avoid the formation of an energetically unfavorable interface.  

\begin{figure}[t!]
    \centering
    \includegraphics[width=0.3\linewidth]{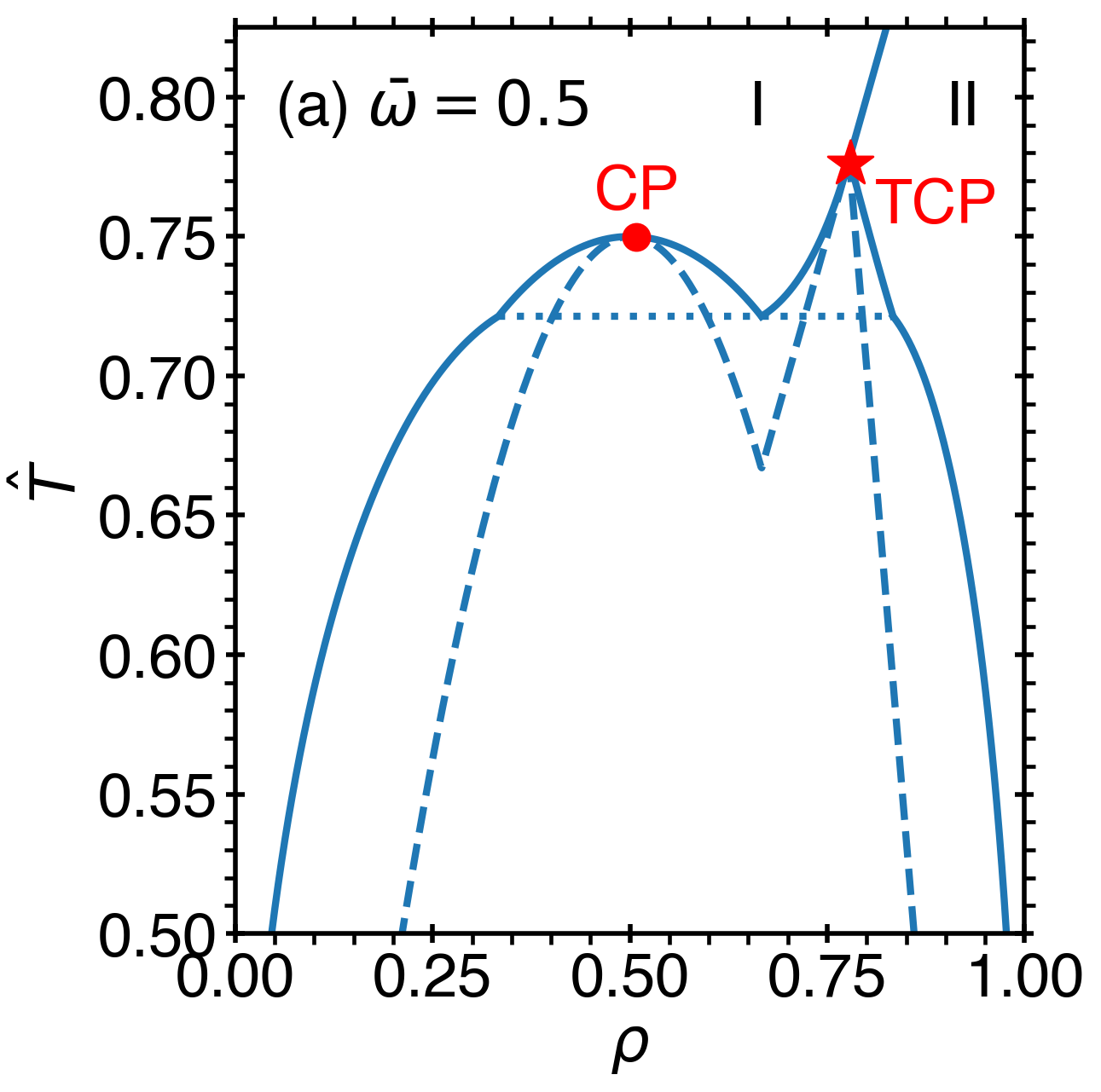}
    \includegraphics[width=0.3\linewidth]{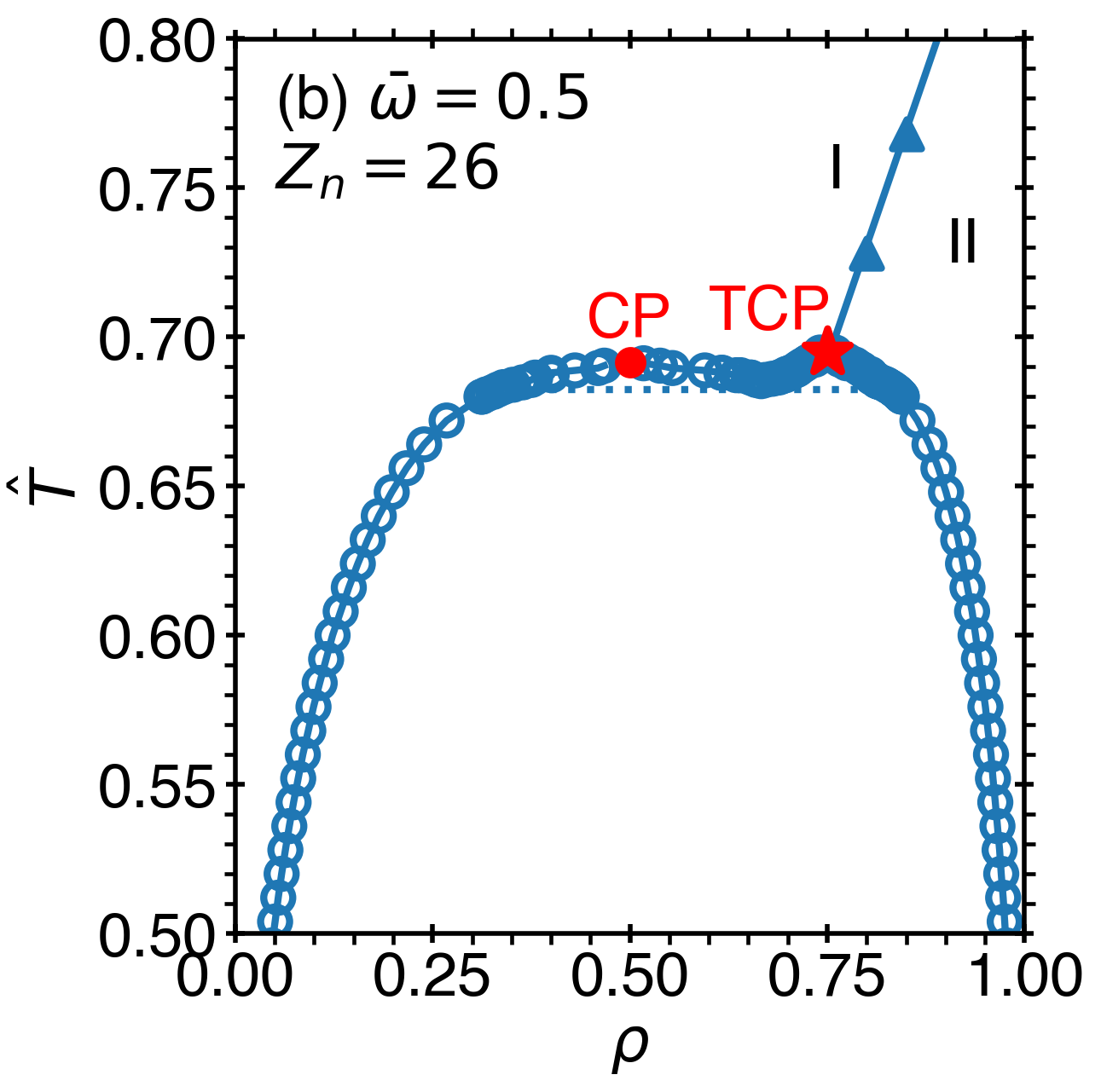}
    \includegraphics[width=0.3\linewidth]{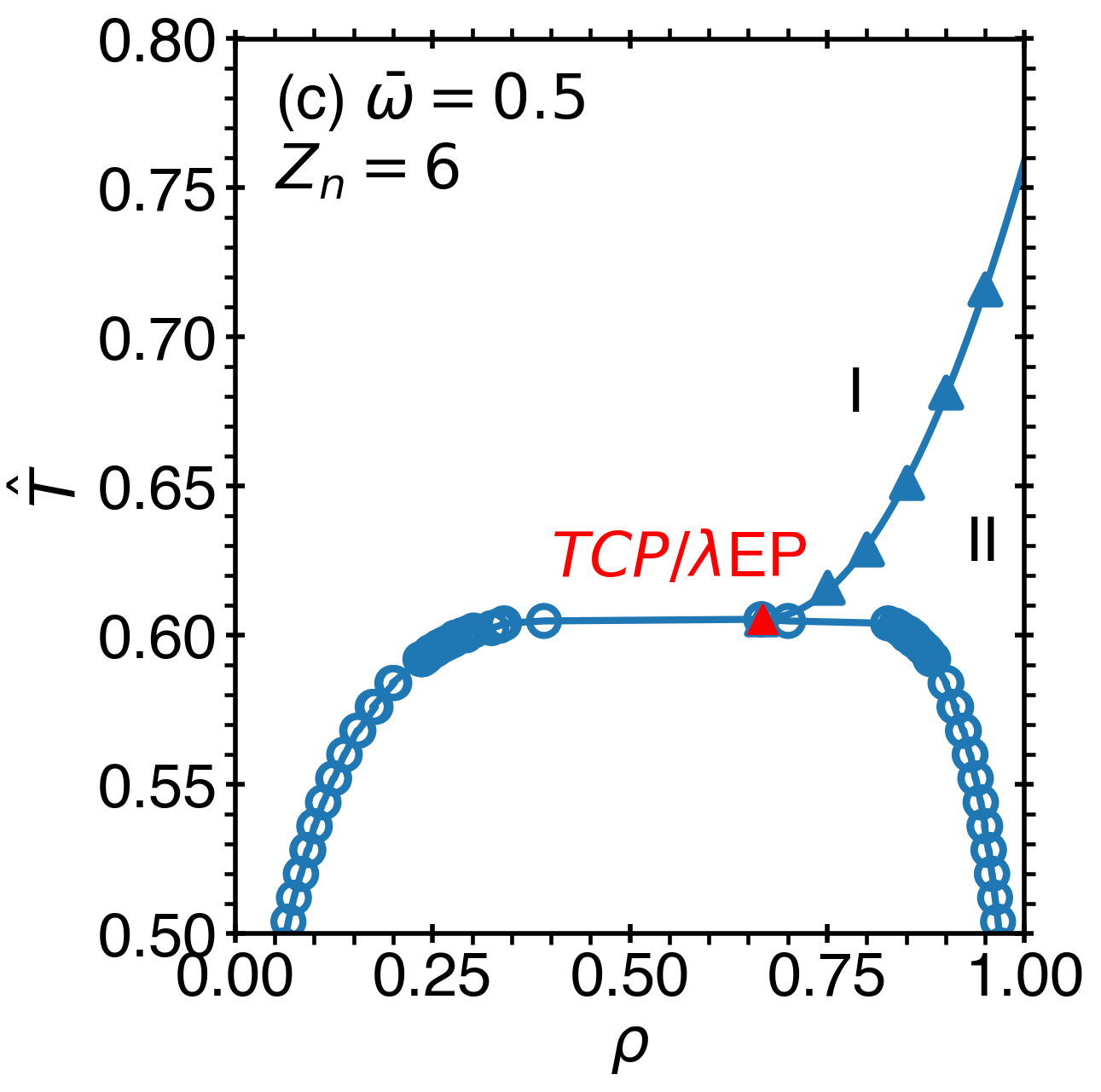}
    \caption{Type II behavior. Comparison of (a) meanfield, (b) MC simulations with coordination number ($Z_n=26$), and (c) MC simulations with $Z_n=6$ for the system with $\bar{\omega}=0.5$. The solid curves represent the liquid-vapor coexistence, the dashed curves the spinodal, and the dotted line the liquid-liquid-vapor triple line, the open circles MC coexistence data, and the triangles MC lambda line data. The red symbols represent: the liquid-gas critical point (circle), the tricritical point (star), and the $\lambda$-end point (triangle).}
    \label{Fig_w0p5}
\end{figure}

In the MC simulations, similar phase behavior is observed. For the system with $\bar{\omega}=0.45$, as illustrated in Fig.~\ref{Fig_App_wBar_0p5}c, a liquid-gas critical point is observed along with an indentation in the liquid branch of liquid-vapor coexistence, while for the system with $\bar{\omega}=0.52$, a tricritical point is observed. However, for both the system with $\bar{\omega}=0.45$ and $\bar{\omega}=0.52$, one only observes either a liquid-gas critical point or a tricritical point. One may hypothesizes that the apparent difference between the MC simulations and the meanfield calculations could be due to the restrictions of the nearest-neighbor approach (with coordination number $Z_n=6$). To verify this hypothesis, we investigated the system with $\bar{\omega}=0.5$ via meanfield calculations, MC simulations with $Z_n=6$, and MC simulations with $Z_n=26$ (including next-nearest neighbors). As presented in Fig.~\ref{Fig_w0p5}, the meanfield behavior is confirmed in the MC simulations with $Z_n=26$, in which a liquid-gas critical point and a tricritical point are observed. For the case of the MC simulations with $Z_n=6$, a tangential merging of the lambda line with the vapor branch of coexistence is shown, at which the system exhibits characteristics of both criticality and tricriticality. 

The discrepancy between the meanfield calculations and the MC simulations with $Z_n=6$ for the system with $\bar{\omega}=0.5$ could be explained from an analogy of the degenerate BC model to an incompressible symmetric ternary mixture, in which interaction between like particles is one ($\omega_{ii}=1$), while the interaction between unlike particles is $\omega_{ij}=\omega_{ii}/2$. This symmetric ternary mixture has an effective $\bar{\omega}_\text{eff} = 1 -\sum\omega_{ij}/\sum\omega_{ii}= 0.5$. One can find that the energy change of both models, at any MC step (Glauber and Kawasaki), are always identical if $Z_n=6$. In contrast, if $Z_n=26$ in a Kawasaki step, when swapped particles occupy lattice sites within the interaction range, the energy change of these models are not equal. Note that in the symmetric ternary mixture with these specific interactions, the critical partial density of each component is $1/3$. If one could identify one component of the ternary mixture with the empty spots of the BC lattice, one would find a critical density of $\rho_\text{c}=2/3$. One observes that, in the degenerate BC model with $Z_n=6$, the liquid-gas critical density and the $\lambda$-end point coincide at $\rho=2/3$, as shown in Fig.~\ref{Fig_w0p5}c.




The effect of increasing the coordination number is reminiscent of lattice discretization considered in other studies~\cite{Panagiotopoulos_Lattice_1999,Moghaddam_Discretization_2005}. Moreover, accurate simulations of the Ising model with varying coordination numbers demonstrate a crossover from fluctuation-induced to meanfield critical behavior in the limit $Z_n\to\infty$~\cite{Luijten_Medium_1996,Luijten_Nature_1998,Anisimov_Crossover_1999,kim_crossover_2003}. Consequently, in a future work, it would be interesting to simulate the BC model for larger values of $Z_n$ to demonstrate this crossover.

\subsubsection{Type III Behavior}

\begin{figure}[t!]
    \centering
    \includegraphics[width=0.3\linewidth]{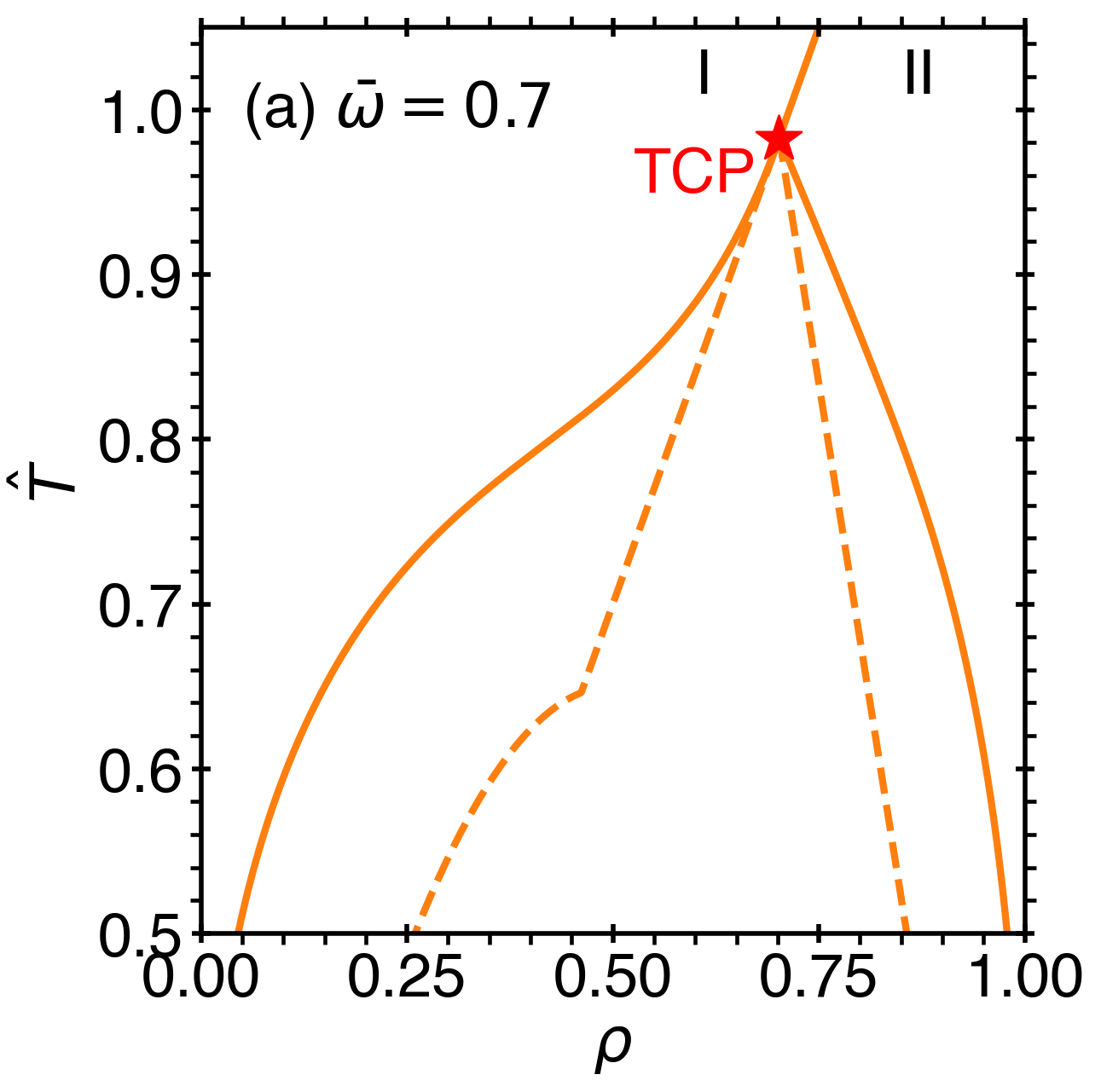}
    \includegraphics[width=0.3\linewidth]{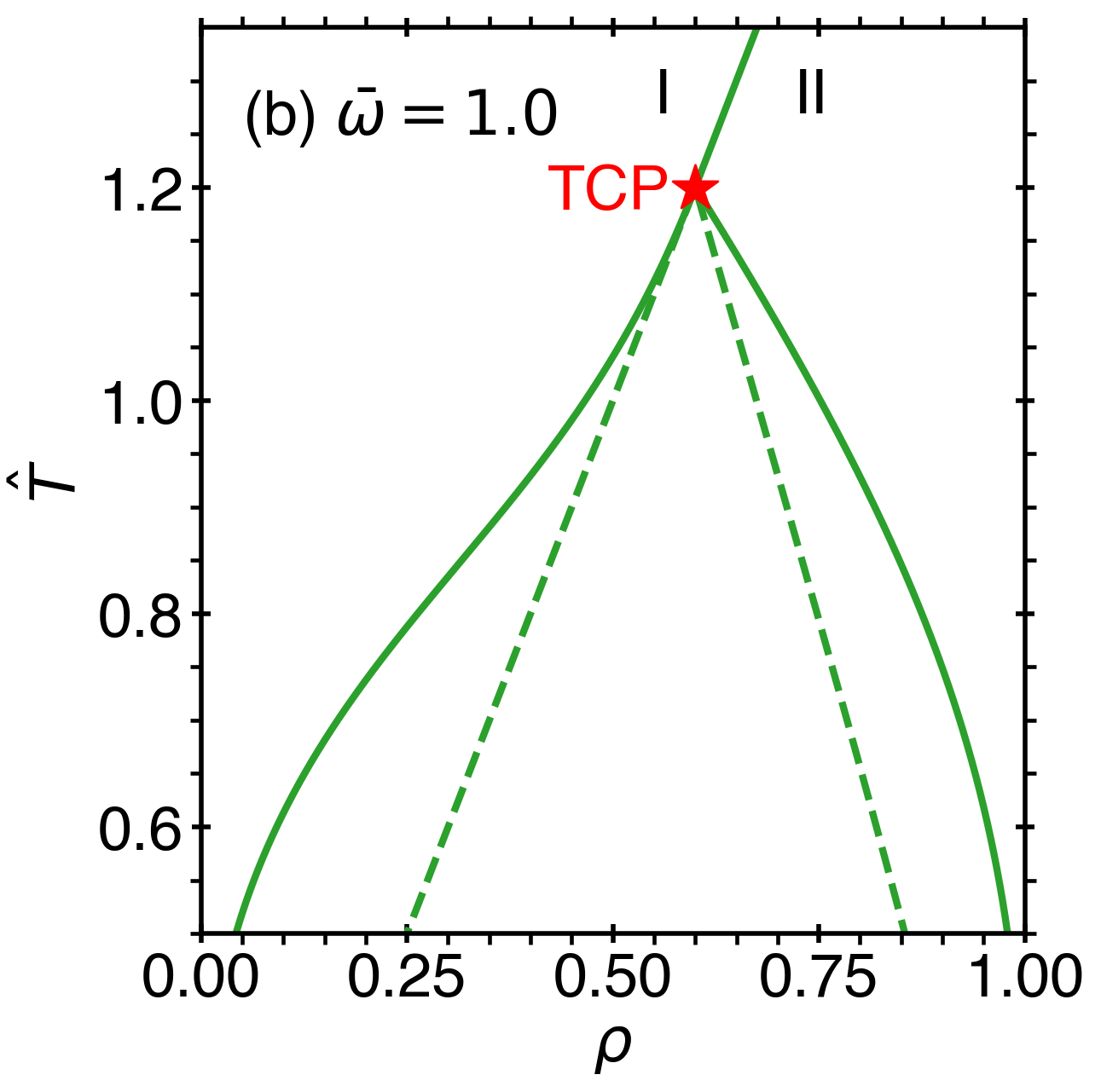}
    \includegraphics[width=0.3\linewidth]{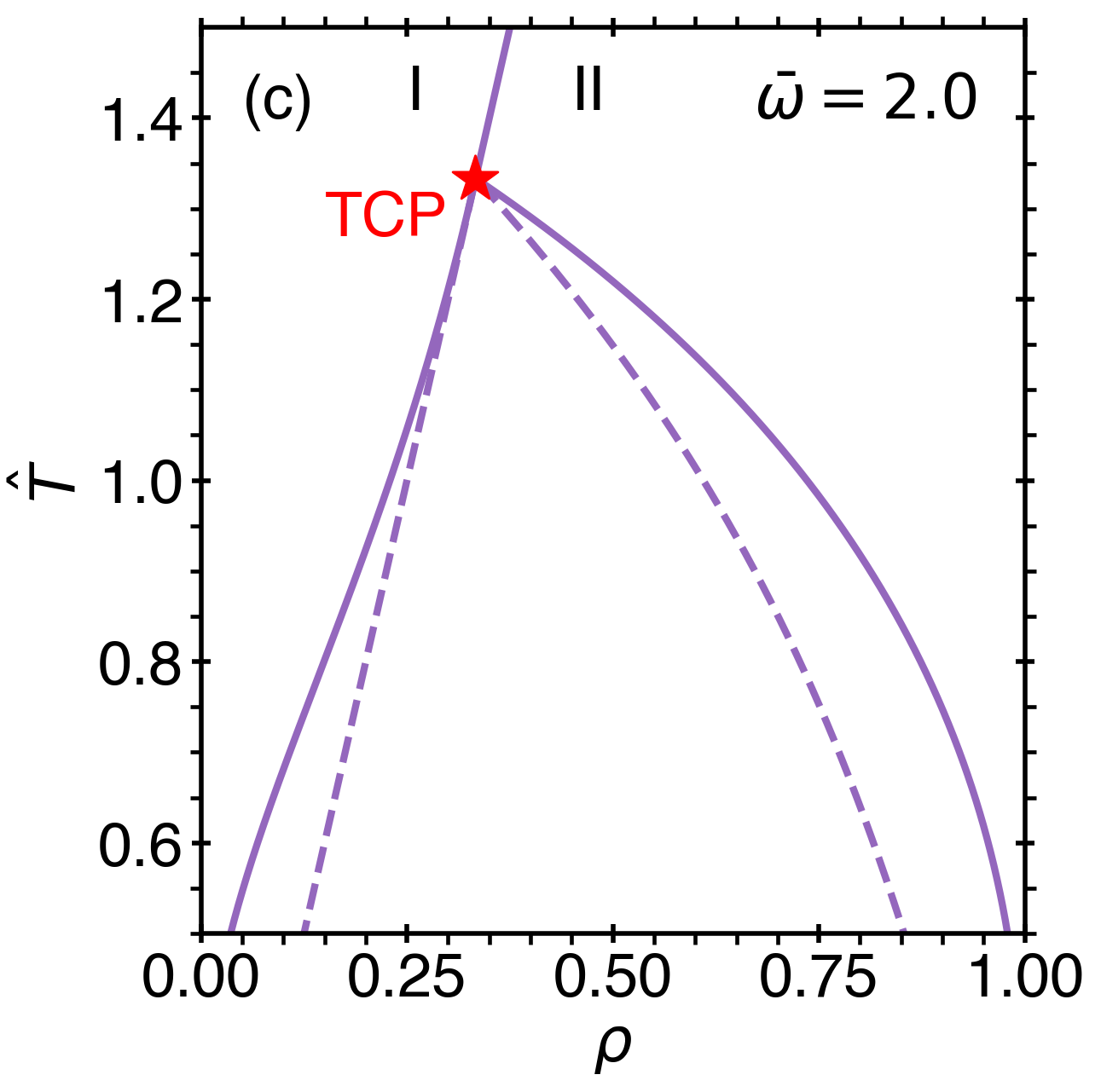}
    \includegraphics[width=0.3\linewidth]{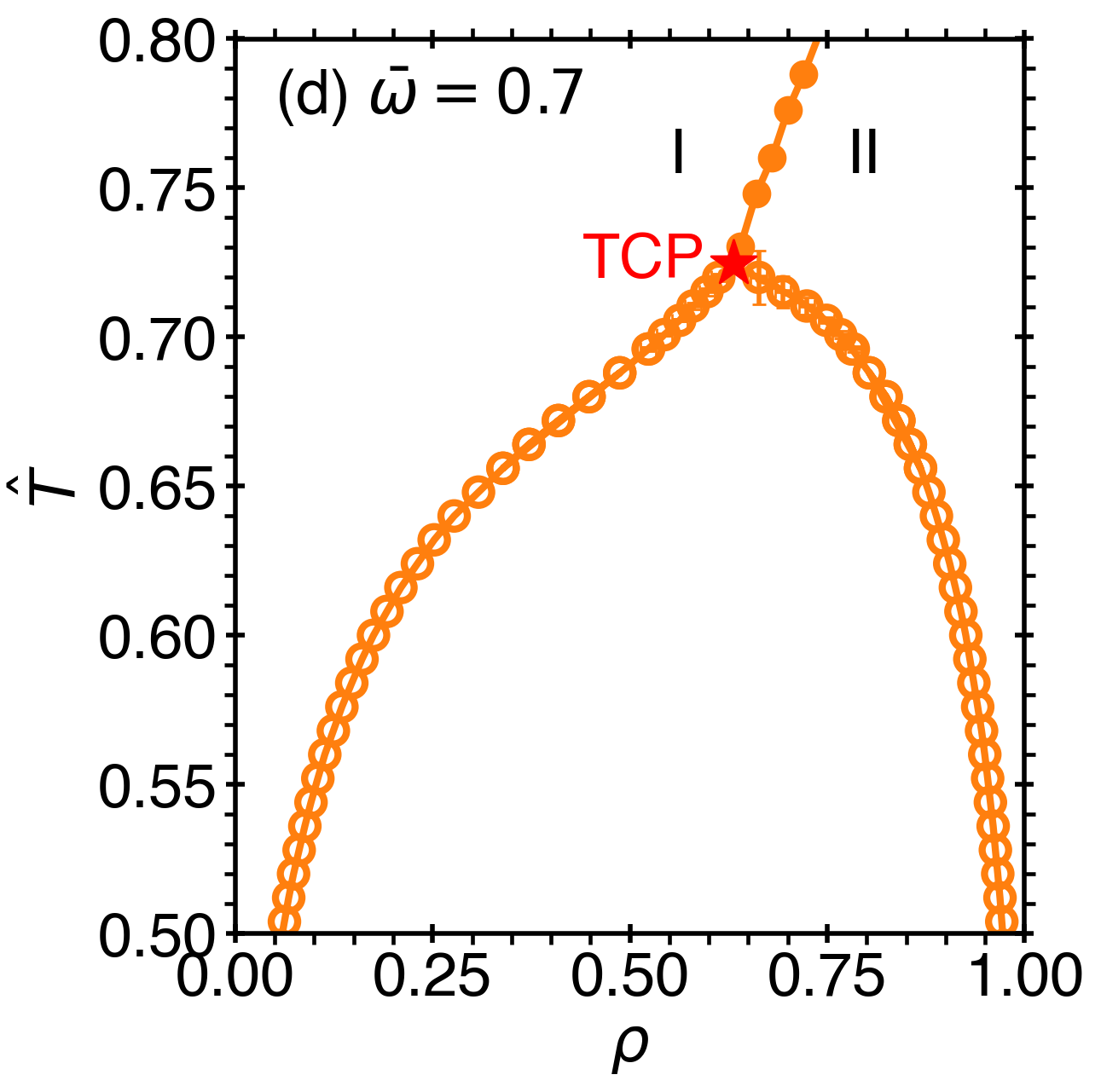}
    \includegraphics[width=0.3\linewidth]{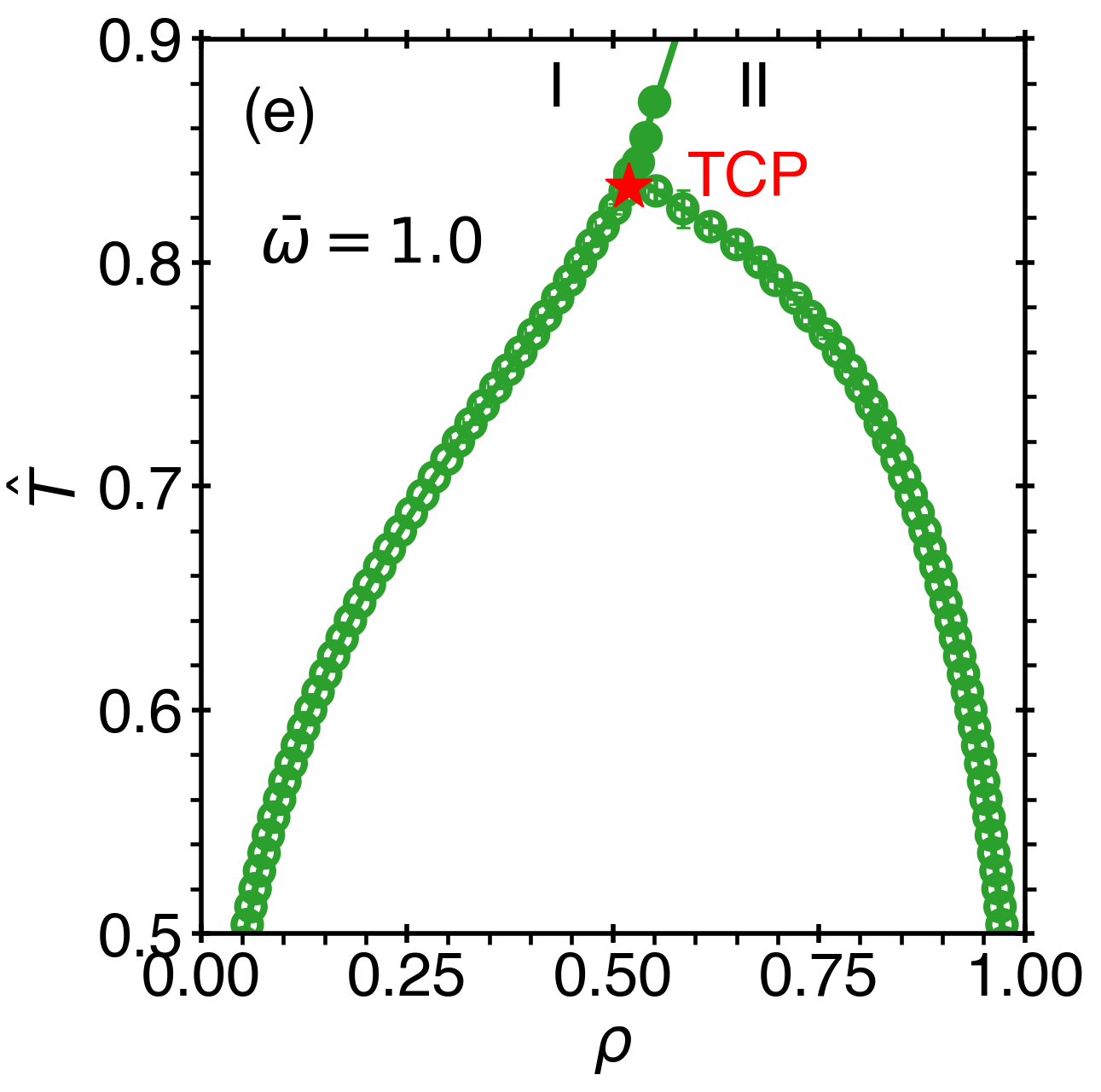}
    \includegraphics[width=0.3\linewidth]{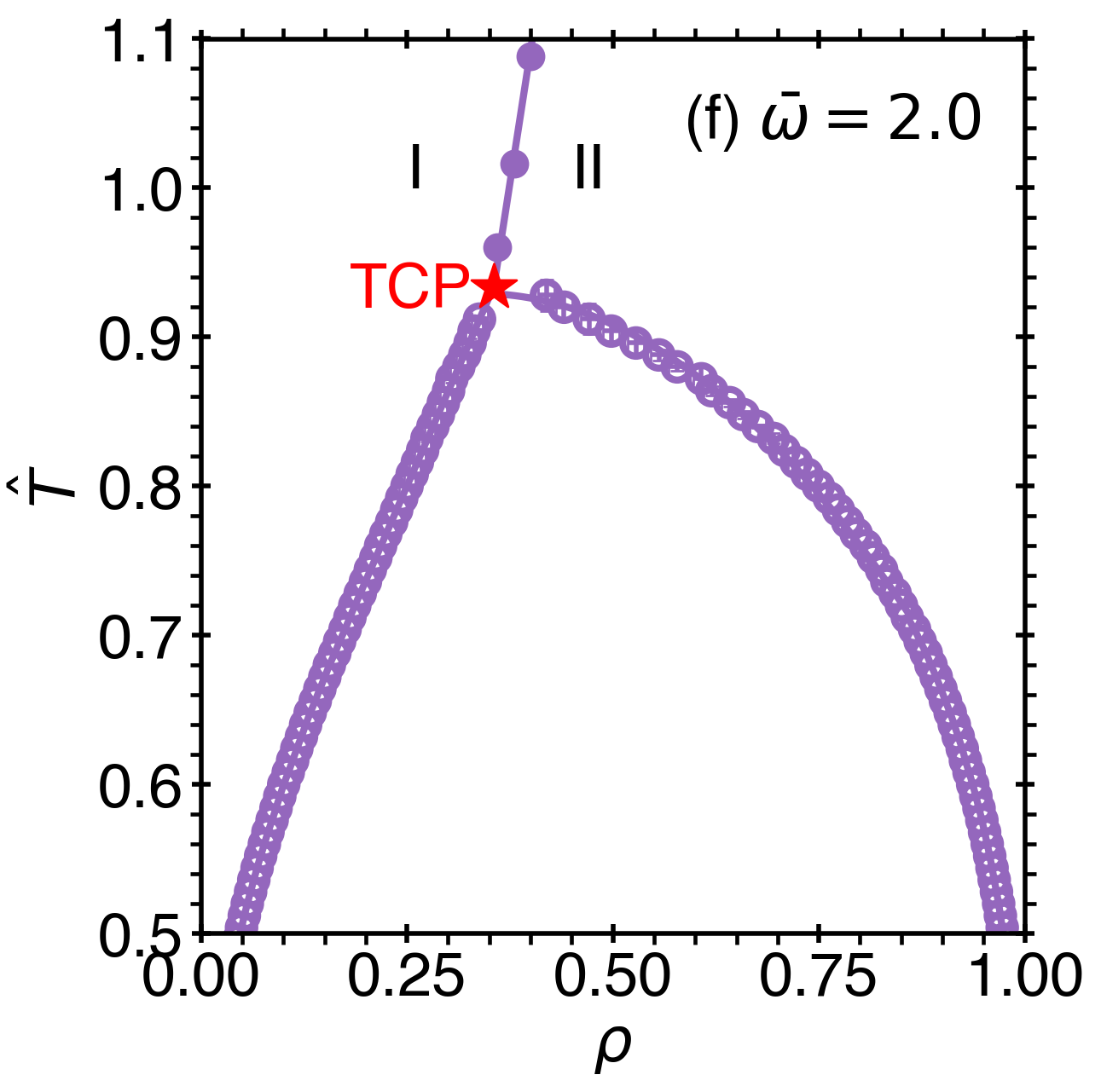}
    \caption{Type III behavior. Comparison of meanfield and MC (with $Z_n=6$) phase diagrams $\bar{\omega}=0.7$ (a,d), $\bar{\omega}=1.0$ (b,e), $\bar{\omega}=2.0$ (c,f) in the degenerate BC model. For meanfield, the solid curves correspond to liquid-gas coexistence, while the dashed curves represent the stability limit (spinodal). For MC, the open circles correspond to the liquid-gas coexistence, the closed circles illustrate the lambda line, while the solid curves represent guidelines for the coexistence. The area marked as ``I'' corresponds to the ``disordered'' fluid, while the area marked as ``II'' corresponds to the ``ordered'' fluid where interconversion depends on temperature and density. The line that separates these two regions is the ``lambda'' line. The lambda line crosses the maximum of the coexistence at the tricritical point (TCP).}
    \label{Fig_MFPhaseDiagrams}
\end{figure}

Type III behavior occurs in the interval $0.538\le \bar{\omega} \le 2.450$, and is characterized by a system that only exhibits a tricritical point. Fig.~\ref{Fig_MFPhaseDiagrams} illustrates a confirmation of this archetype in the degenerate BC model, obtained from meanfield calculations and MC simulations, for three values of the normalized nonideality parameters, $\bar{\omega}$, in this interval. As demonstrated in Fig.~\ref{Fig_MFPhaseDiagrams}(a-c), the lambda line arrives at the top of the liquid-gas coexistence --  where the liquid and gas branches meet at a sharp angle. The geometry is characteristic of a symmetrical TCP, at which second-order transitions between disordered (I) and ordered (II) fluid meets the line of the liquid-gas first order phase transition. For example, such geometry of the phase diagram is observed in the liquid mixture of helium isotopes (see Fig.~\ref{Fig_HePhaseDiagram}b), polymer solutions at an infinite degree of polymerization near the theta point~\cite{Anisimov_PolymerTricrit_2024,Anisimov_Mesoscopic_2024}, and in the lattice restricted primitive model of ionic solutions~\cite{Panagiotopoulos_Lattice_1999,Panagiotopoulos_Tricriticality_2003}. 

Figure~\ref{Fig_MFPhaseDiagrams}(d-f) depicts the BC model, obtained by MC simulations, for the same values of $\bar{\omega}$ shown in (a-c). Qualitatively, these results agree with the meanfield calculations. The location of the TCP was obtained from the intersection of linear fits of the gaseous branch of the equilibrium density and the lambda line simulation data. As expected, the values of the critical and tricritical temperatures are systematically lower than those obtained in the meanfield approximation. This is a well-known effect of critical fluctuations~\cite{Anisimov_Mesoscopic_2024}. It is well-known that the tricritical point exhibits meanfield behavior with logarithmic corrections caused by fluctuations~\cite{Lawrie_Tricritical_1984,Knobler_Tricritical_1984}. Within the accuracy of our MC simulation data, however, the curve through the liquid branch is a linear function with an empirical parabolic correction.  

\subsubsection{Type IV Behavior}
\begin{figure}[t!]
    \centering
    \includegraphics[width=0.3\linewidth]{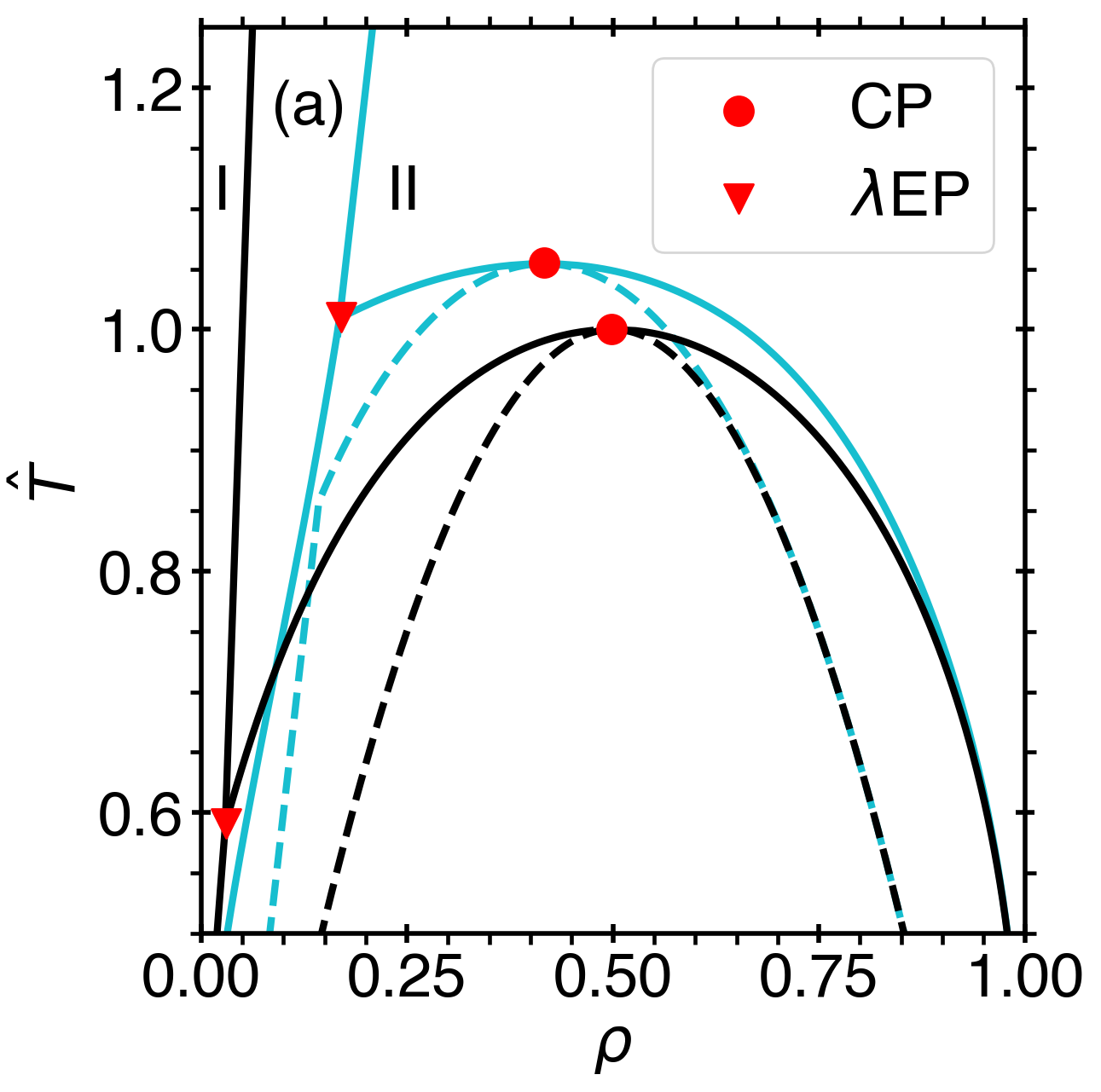}
    \includegraphics[width=0.3\linewidth]{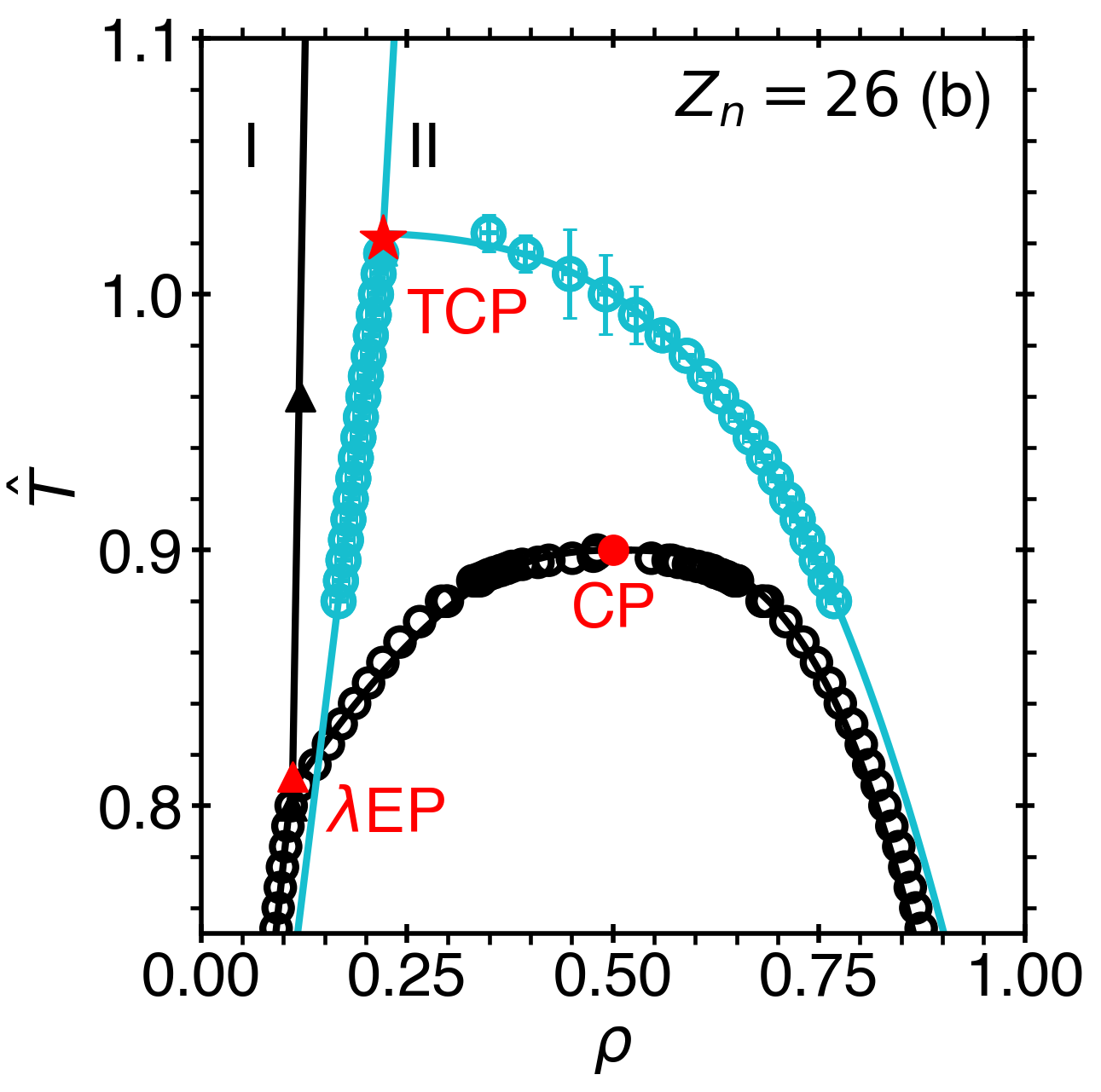}
    \includegraphics[width=0.3\linewidth]{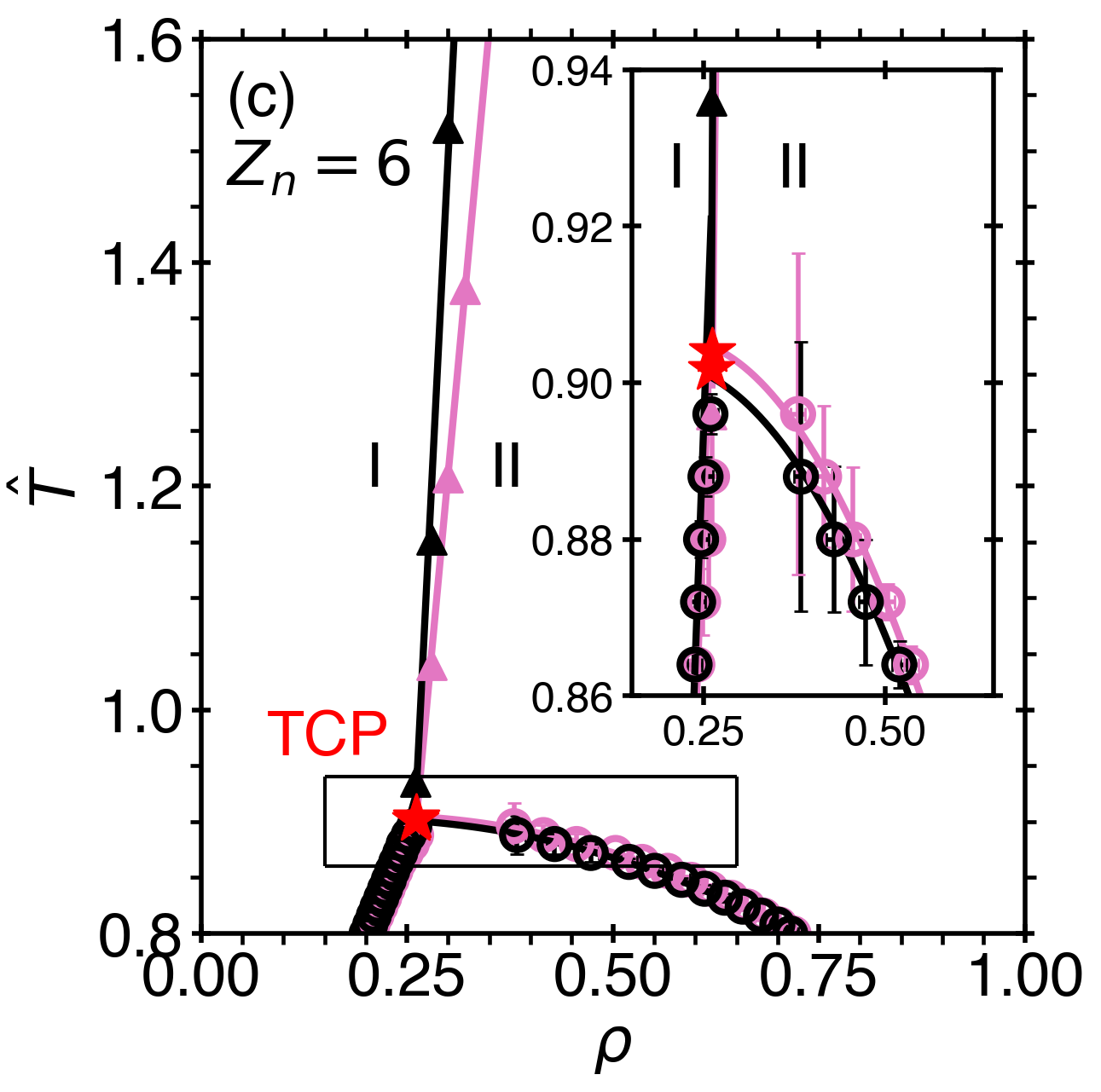}
    \caption{Type IV behavior. Temperature-density phase diagrams for large values of $\bar{\omega}$ as obtained in (a) meanfield calculations (b) MC simulations with $Z_n=26$, and (c) MC simulations with $Z_n=6$. In (a,b), the colors correspond to $\bar{\omega}=3$ (cyan) and $\bar{\omega}=10$ (black), while in (c), the curves correspond to $\bar{\omega}=5$ (pink) and $\bar{\omega}=10$ (black). The red circles, triangles, and stars correspond to the locations of the critical point, lambda end point, and tricritical point, correspondingly. In (c), the inset shows a zoomed region near the tricritical points (almost overlapping for the two cases), indicated by the boxed region.}
    \label{Fig_LargeWbarComp}
\end{figure}

Type IV behavior occurs for systems where $\bar{\omega}\ge 2.450$, and is characterized by a LGCP with a lambda line that intersects the vapor branch of liquid-gas coexistence. Figure~\ref{Fig_LargeWbarComp} shows the phase behavior of the BC model for very large values of $\bar{\omega}$ obtained from meanfield calculations and MC simulations. One finds that tricriticality disappears and a critical end point emerges in the meanfield prediction. The crossing of the lambda line with the coexistence on the vapor branch indicates the formation of two gaseous phases, a disordered gas (I) at $\hat{T}<\hat{T}_\lambda$ and an ordered gas (II) at $\hat{T}>\hat{T}_\lambda$.

At the $\lambda$-end point, a critical mixture of disordered-ordered gas is in coexistence with ordered liquid (II). For $\hat{T}_\lambda\le \hat{T}\le\hat{T}_\text{CP}$, the ordered gas is in coexistence with ordered liquid. It should be noted that in the degenerate BC model, the ordered liquid is comprised of either liquid-1 or liquid-2, as phase amplification causes one phase to grow at the expense of the other. This is different from the quadruple phase coexistence predicted by Furman et al.~\cite{Furman_Global_1977} in the binary non-interconverting system, in which four phases could coexist: two ordered vapors with complementary fractions ($x_V$ and $1-x_V$) and two ordered liquids with complementary fractions ($x_\ell$ and $1-x_\ell$). This apparent violation of Gibbs' phase rule for a binary fluid stems from the degeneracy of the coexistence equation (see Refs.~\cite{Meijering_TernaryI_1950,Meijering_TernaryII_1951}). Therefore, phase amplification suppresses the alternative liquid and vapor, leaving only the coexistence between one ordered vapor and one ordered liquid. This occurs over a range of temperatures, in accordance with Gibbs' rule of phases for a single-component substance.

Similar to what was observed in Type II behavior, MC simulations only exhibit Type IV behavior for $Z_n=26$ as shown in Fig.~\ref{Fig_LargeWbarComp}b. Note that, within the accuracy of the MC simulations, for the system with $\bar{\omega}=3$ and $Z_n=26$, a tricritical point is observed with a nearly horizontal liquid branch of the liquid-vapor coexistence. This system may be one of the last systems of Type III behavior or the onset of Type IV behavior in MC simulations. However, for the system with $\bar{\omega}=10$ and $Z_n=26$, the order-to-disorder transition in the gaseous phase accompanied by an interception of the $\lambda$-line with the gaseous branch of the liquid-vapor coexistence is observed in agreement with the meanfield calculations

As presented in Fig.~\ref{Fig_LargeWbarComp}c, there is a significant difference between the meanfield and MC simulation results for the degenerate BC model at high values of $\bar{\omega}$ and $Z_n=6$. The MC simulations demonstrate that the ordered gaseous phase is never formed, and that the tricritical points collapse to the same temperature ($\hat{T}\to 0.9$) and density ($\rho\to 0.25$) in the limit of $\bar{\omega}\to\infty$. The discrepancy between the meanfield calculations and the MC simulations is due to the specifics of the lattice model with short range of interactions (with coordination number, $Z_n=6$). For example, when $\bar{\omega}\to\infty$, then $\omega_{12}\to -\infty$, and hence, the Metropolis factor employed in the MC simulations for nearest-neighbor lattice sites becomes of the order $\sim\exp(\omega_{12}/T)$, which is negligible above absolute zero temperature, such that particles of type 1 and 2 will never be allowed to occupy neighboring sites. This lattice is restricted such that the interaction ranges of unlike particles should not intersect for $Z_n=6$. Meanfield predicts that the transition to the Type IV archetype is achieved when the critical density is approximately $1/4$, but at such a low density and short range of interactions, the entropy of the disordered gas is higher than the entropy of the ordered gas; thus, the transition from ordered to disordered gas is forbidden in the MC simulations.



\subsection{$P$-$\rho$ and $P$-$T$ phase diagrams}~\label{Sec_Results_PT_Data}
\begin{figure}[b!]
    \centering
    \includegraphics[width=0.45\linewidth]{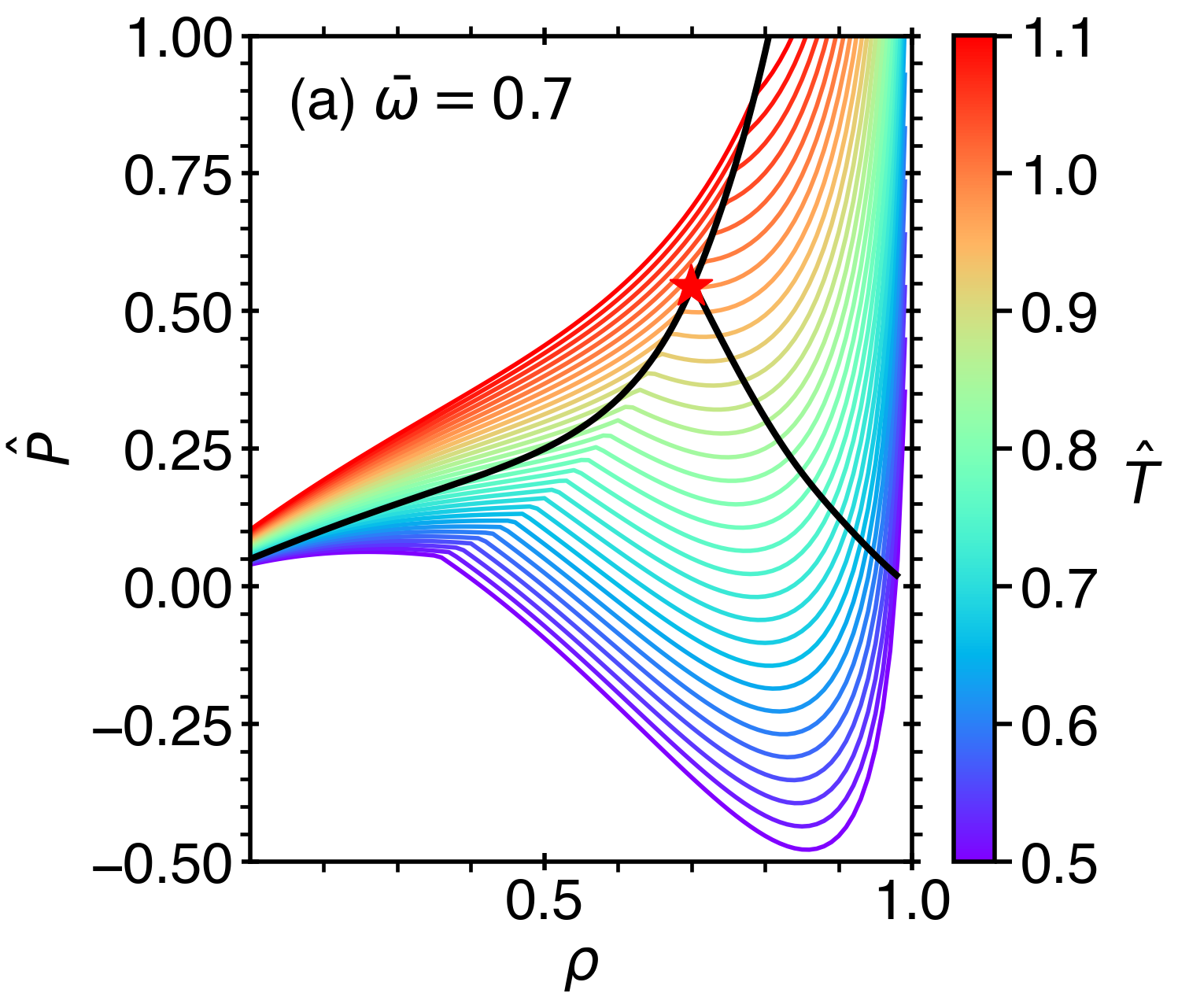}
    \includegraphics[width=0.4\linewidth]{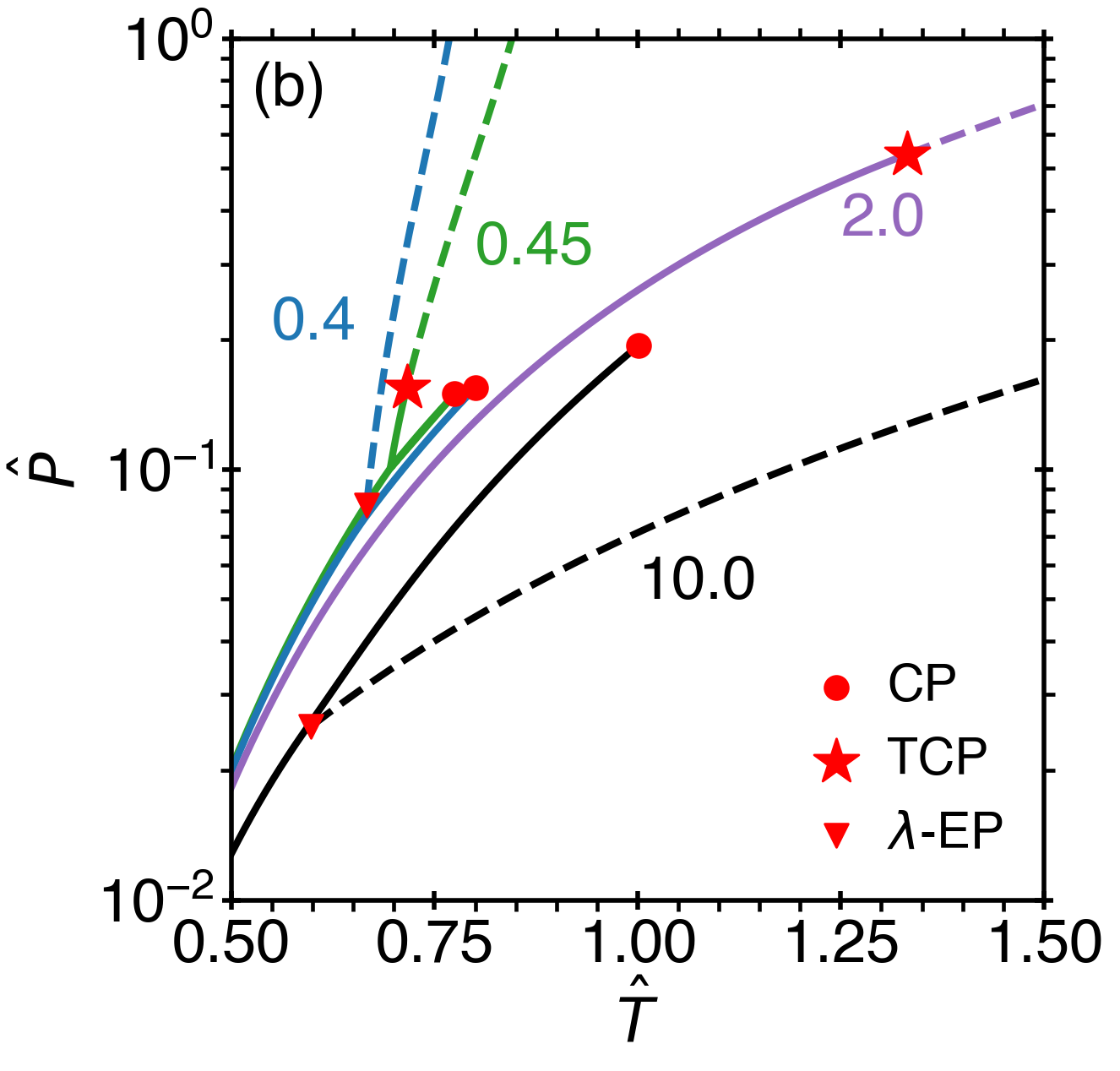}
    \caption{Meanfield pressure-density (a) and pressure-temperature (b) phase diagrams for the degenerate BC model. In (a), the calculations are shown for various isotherms (colored curves) in the system with $\bar{\omega}=0.7$. The black curves depict coexistence and the star indicates the tricritical point. In (b), pressure-temperature calculations along liquid-gas coexistence are presented for a representative system exhibiting each archetypical behavior. The dashed curves illustrate the behavior of the lambda lines for each case.}
    \label{Fig_Pressure}
\end{figure}
In this section, pressure calculations obtained in the meanfield approximation are presented to illustrate the behavior of the four different archetypes. Figure~\ref{Fig_Pressure}a shows a typical pressure-density phase diagram for $\bar{\omega}=0.7$ exhibiting type III behavior, in which the pressure was calculated from Eq.~(\ref{Eq_MF_pressure}). The cusp in the pressure-temperature isotherms reveal the location of the lambda line above the liquid-gas coexistence. In addition, a system representing each archetype is presented in pressure-temperature space in Fig.~\ref{Fig_Pressure}b. For $\bar{\omega}=0.4$ and $\bar{\omega}=10$, the line of liquid-liquid critical points terminates at the $\lambda$-end point below the liquid-gas critical point (similar to the $\lambda$-end point in $^4$He at the saturated-vapor pressure -- see Fig.~\ref{Fig_HePhaseDiagram}a). Note that the lambda line approaches the coexistence from a higher pressure when it crosses the liquid branch, while it crosses the coexistence from a lower pressure when it crosses the vapor branch. For systems with $\bar{\omega}=0.45$ and $2.0$, at the tricritical points, a smooth transition in $P$-$T$ coordinates from the liquid-gas coexistence to the lambda line is observed, in which the two curves meet with the same slope.

\section{Discussion}~\label{Sec_Discussion}
This Section is separated into two parts. In Section~\ref{Sec_Discussion_Nature}, the nature of the tricritical point is elaborated with respect to other tricritical systems, while in Section~\ref{Sec_Discussion_Comparison}, the critical behavior of the BC model is presented across all four archetypes. 

\subsection{Nature of Symmetrical Tricriticality in the BC Model}~\label{Sec_Discussion_Nature}
\begin{figure}[t!]
    \centering
    \includegraphics[width=0.418\linewidth]{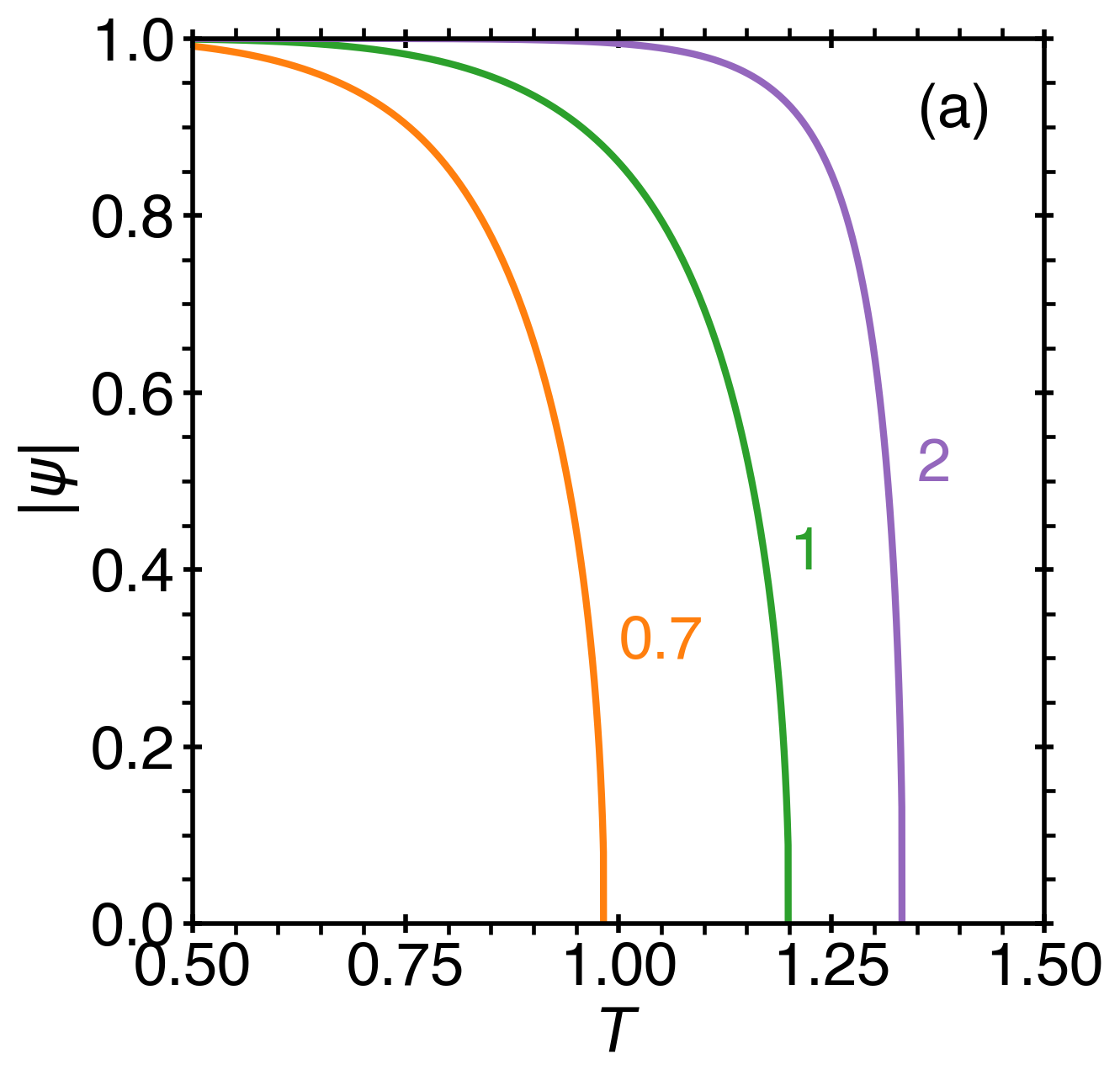}  \includegraphics[width=0.4\linewidth]{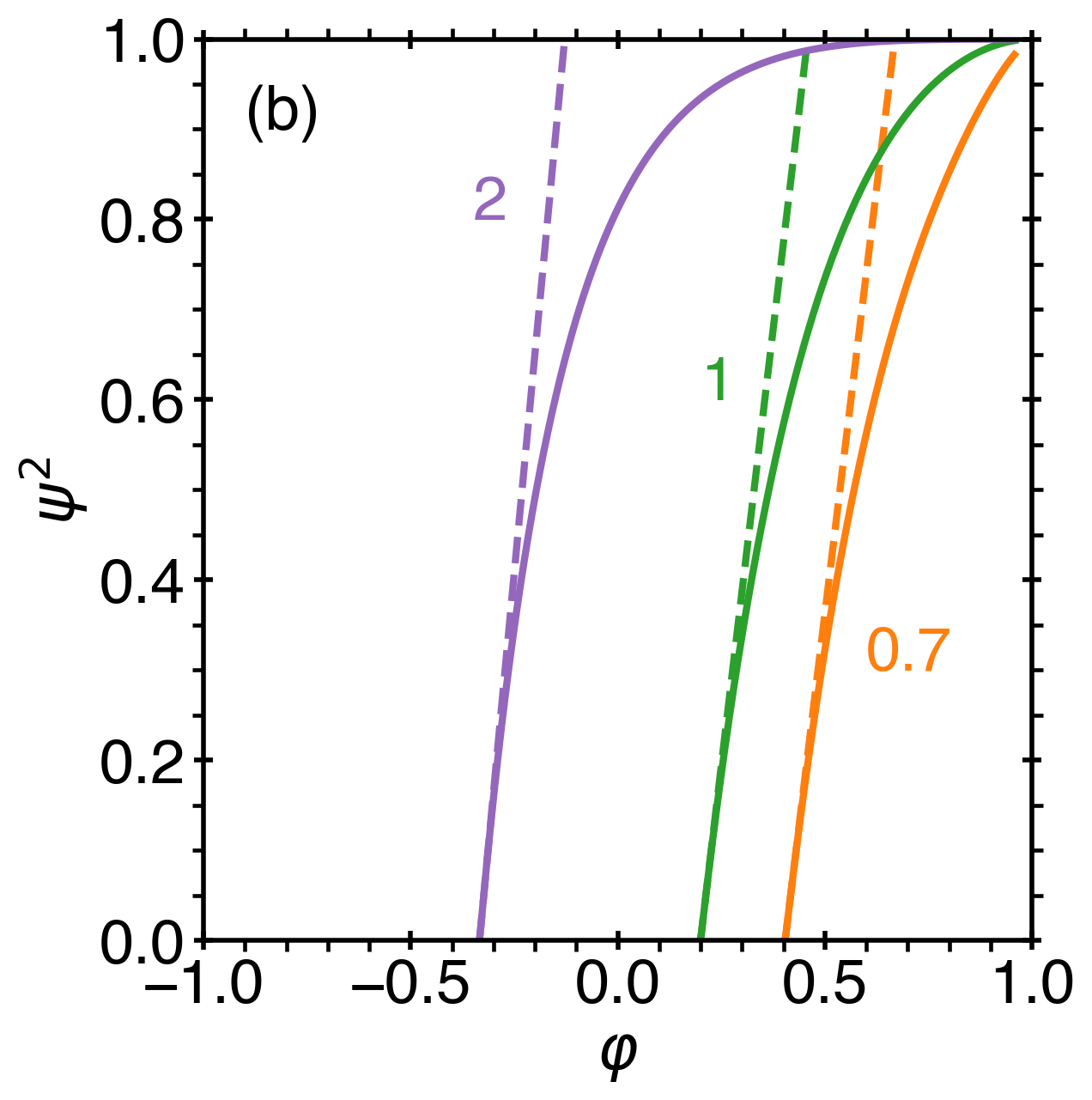}
    \includegraphics[width=0.4\linewidth]{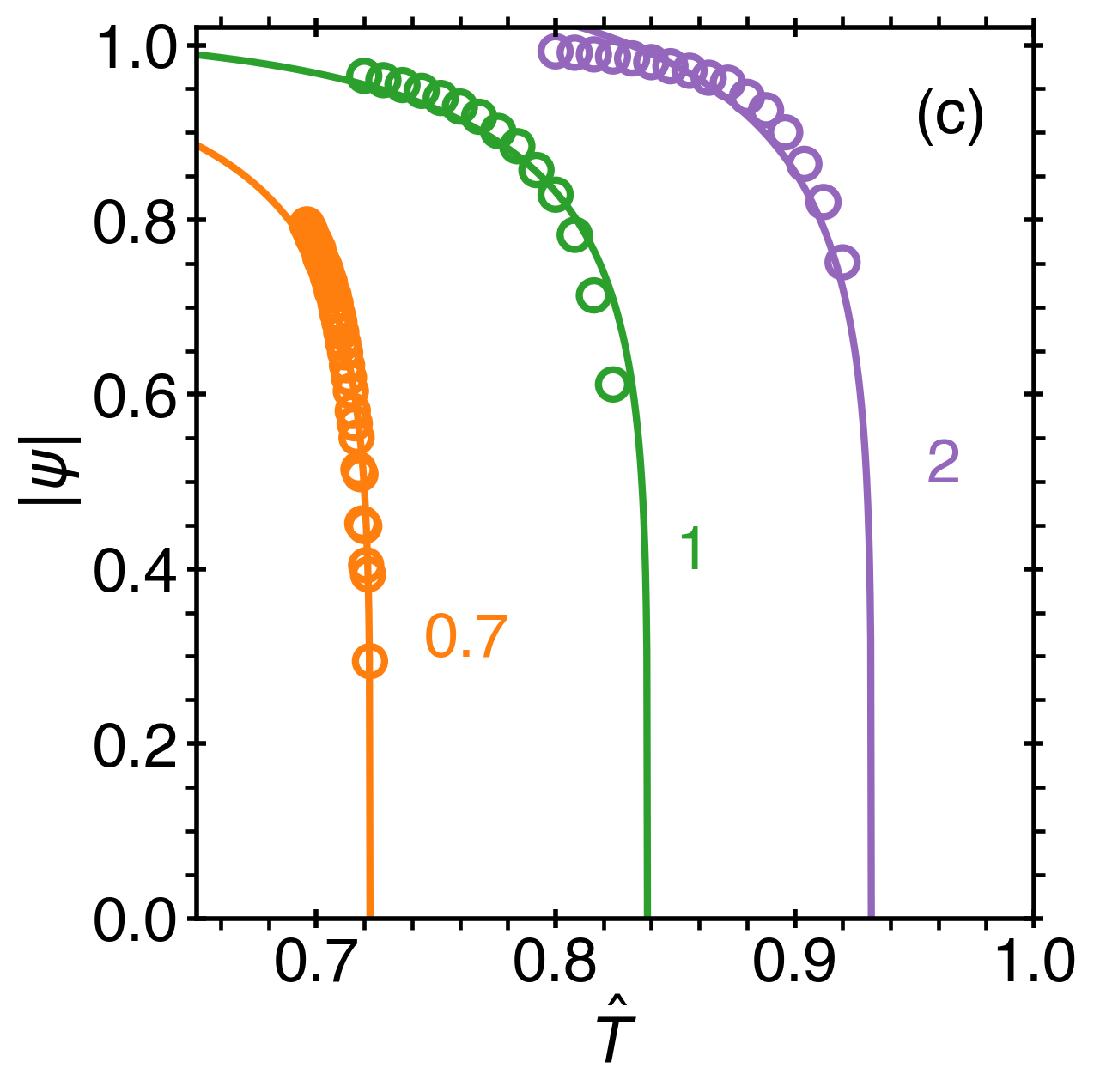}
    \includegraphics[width=0.4\linewidth]{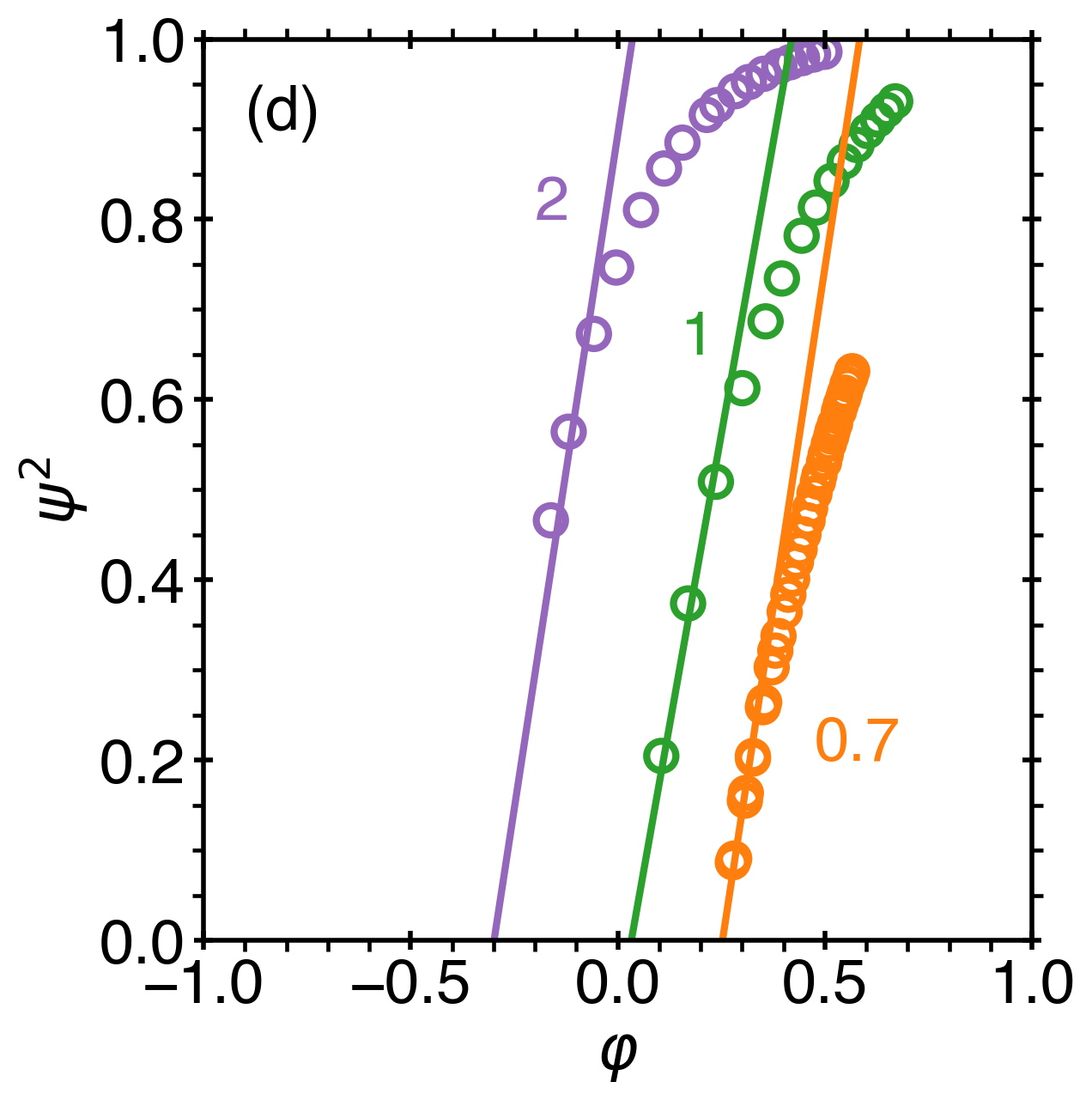}
    \caption{Molecular fraction calculations in meanfield (a,b) and MC simulations (c,d) for $\bar{\omega} = 0.7$ (orange), $1.0$ (green), and $2.0$ (purple). Along the liquid branch of coexistence the nonconserved order parameter, $\psi$, is related to the interconversion fraction as $\psi = 2x-1$, as (a,c) a function of normalized temperature and (b,d) as a function of the conserved order parameter $\varphi$, related to the overall density as $\varphi = 2\rho-1$. The open circles in (c,d) correspond to MC simulation data with $Z_n=6$, while in (a,c), the curves depict the semi-empirical fit to $|\psi|\sim|\Delta\hat{T}|^{1/4} - a|\Delta\hat{T}|^{1/2}$, where $a$ is a constant and $\Delta\hat{T}$ is the reduced distance to the tricritical point. On (b,d) the lines correspond to the asymptotic linear relation between the order parameters in the vicinity of the tricritical point.}
    \label{Fig_comboX}
\end{figure}

Figure~\ref{Fig_comboX}a demonstrates a relationship between the nonconserved order parameter, $\psi$, related to the interconversion fraction as $\psi = 2x-1$, and the conserved order parameter $\varphi$, related to the overall density as $\varphi = 2\rho-1$, in the ordered phase (II). It is predicted that in this phase, asymptotically close to the tricritical point, $\psi^2\sim\varphi$ (see Appendix A). This relationship is observed in other tricritical systems, such as the $^3$He-$^4$He mixture near the transition to superfluidity and near the (tricritical) theta-point of polymer solutions~\cite{Anisimov_PolymerTricrit_2024}. As demonstrated by the MC simulation data presented in Fig.~\ref{Fig_comboX}(c,d), this behavior is also valid for the BC model beyond the meanfield approximation.

\begin{figure}[t!]
    \centering
    \includegraphics[width=0.4\linewidth]{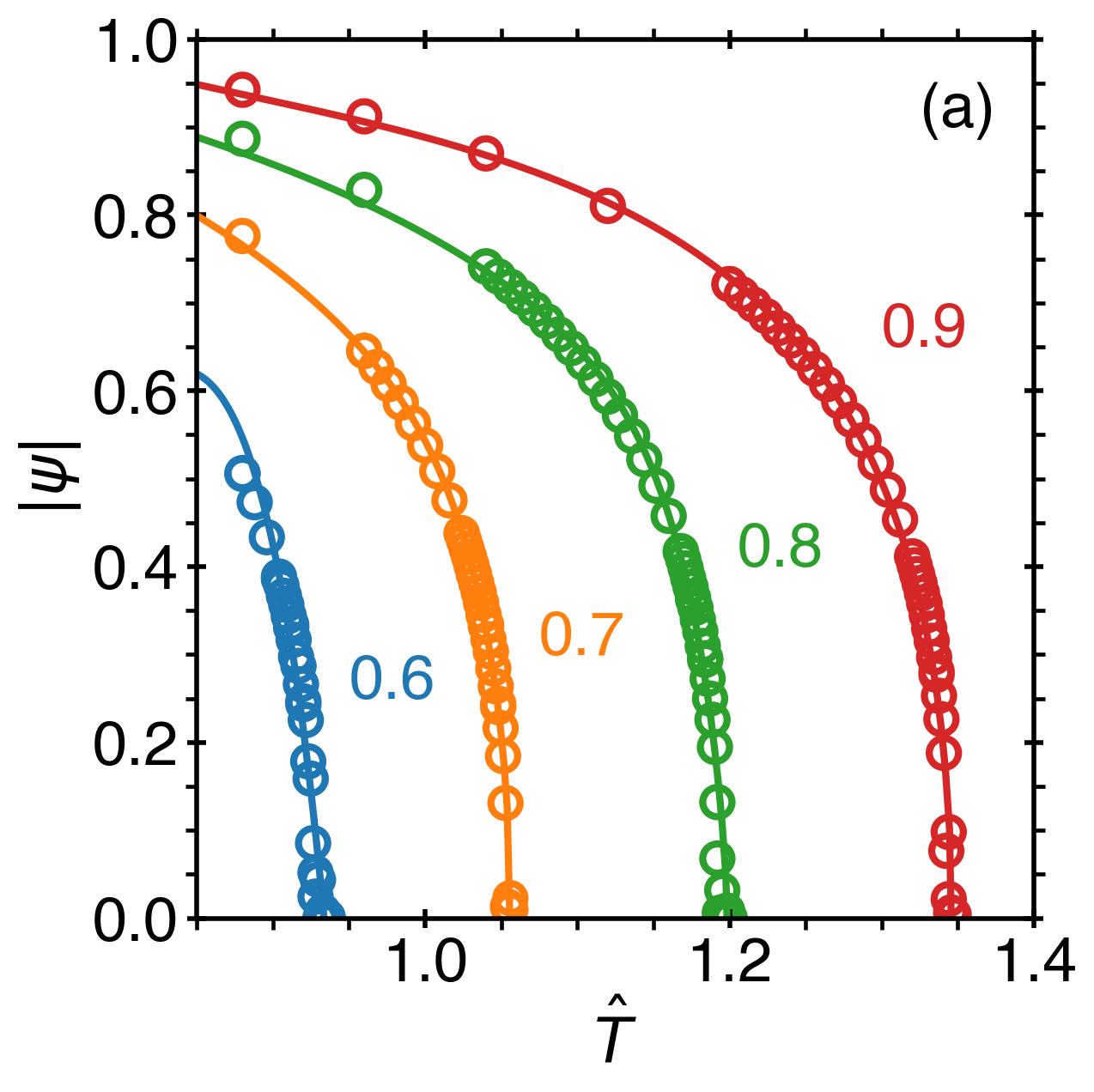}
    \includegraphics[width=0.4\linewidth]{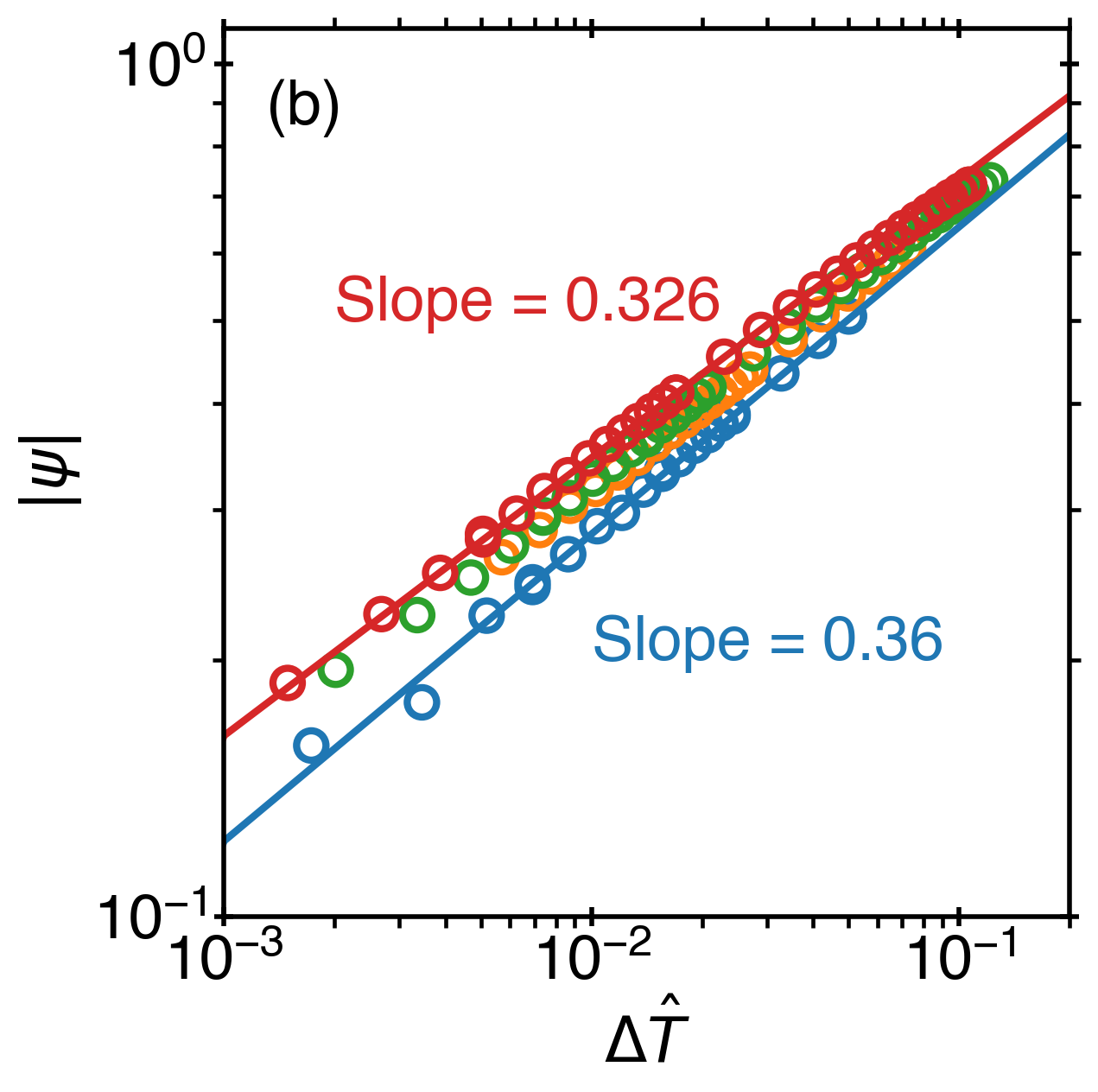}
    \caption{Monte Carlo simulations of the molecular fraction as a function of reduced temperature for $\bar{\omega}=1.0$ with $Z_n=6$ near the lambda transition above the tricritical point for four isochores: $\rho=0.6$ (blue), 0.7 (orange), 0.8 (green), and 0.9 (red). In (a), the solid curves depict the prediction of the scaling theory with symmetric non-asymptotic (Wegner) corrections~\cite{Wang_Asymmetry_2006,Wang_Asymmetry_2007}, while in (b), the prediction of scaling theory that the nonanalytic relation between the fraction and temperature along non-critical isochores scales as $\beta =0.326$ for 3D Ising criticality when $\rho=1.0$. However, the exponent may exhibit a crossover to $\beta^*=0.36$ at smaller densities, which could be attributed to Fisher's renormalization~\cite{Fisher_Hidden_1968} that predicts $\beta^*=\beta/(1-\alpha)$.}
    \label{Fig_MC_xVsT}
\end{figure}

Figure~\ref{Fig_MC_xVsT} shows the behavior of the interconversion order parameter as a function of temperature along various isochores in phase II, obtained by MC simulations for $\bar{\omega}=1$. For a system at overall density $\rho=1.0$, it is expected that far away from the tricritical point, the behavior of the molecular fraction will follow a typical power law for all Ising-like second-order transitions with a critical exponent $\beta \approx  0.326$ in the 3D Ising universality class~\cite{Anisimov_Mesoscopic_2024}. However, as the overall density is decreased and the tricritical point is approached, this critical exponent changes. It could be that this change in the critical exponents is related to Fisher's renormalization~\cite{Fisher_Hidden_1968}, which originates from a nonanalytic relation between isochores (MC path) and isobars (theoretical, Ising-model) and is due to the heat-capacity singularity with exponent $\alpha$~\cite{Anisimov_Mesoscopic_2024}. In phase I (disordered), the value of the nonconserved order parameter, $\psi$, is always zero, being independent of $\varphi$. This feature also supports the Ising-like nature of BC phase transitions, discussed in our earlier publications~\cite{Longo_Interfacial_2023,Buldyrev_BCM_2024}. 

We note that the Landau theory also predicts that the reason for the emergence of symmetrical tricriticality is a coupling between the nonconserved vector-like order parameter, $\psi$, and the conserved order parameter, $\varphi$, as $\lambda\psi^2\varphi$, where $\lambda$ is a coupling constant. A qualitative demonstration, based on the Landau theory of phase transitions, of the emergence of tricriticality that confirms this interpretation in the degenerated BC model, is given in Appendix~\ref{Sec_Appendix_Landau}. The BC model's tricritical phase behavior that most resembles that of a $^3$He-$^4$He mixture, albeit qualitatively, occurs for $\bar{\omega}=2$ with similar values of the normalized critical parameters and slopes of the lambda lines (proportional to $\text{d}T_\lambda/\text{d}\varphi\simeq \bar{\omega}\propto\lambda$).

\subsection{Critical Behavior Spanning All Four Archetypes}~\label{Sec_Discussion_Comparison}
\begin{figure}[t!]
    \centering
    \includegraphics[width=0.5\linewidth]{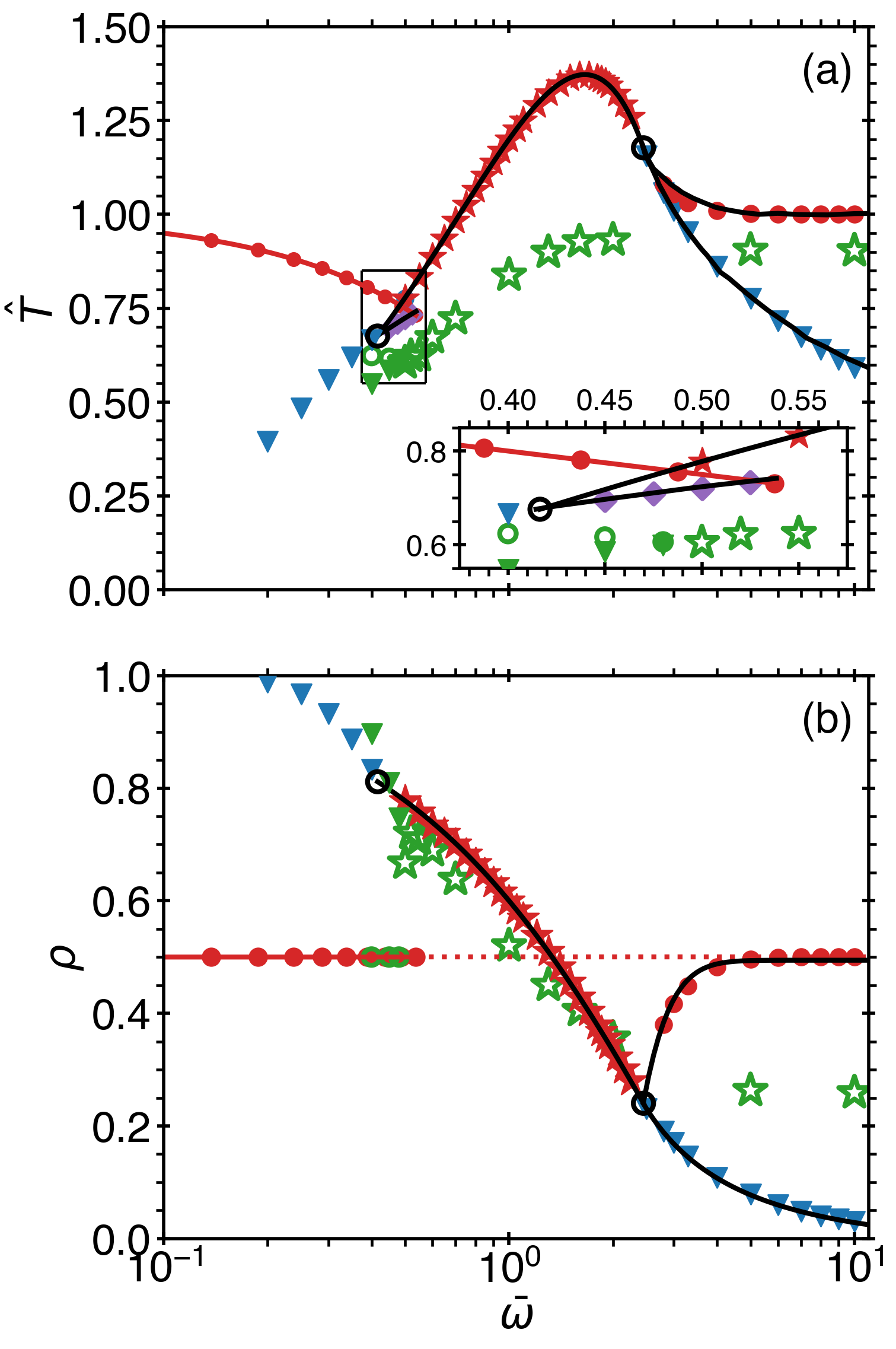}
    \caption{Multicritical behavior of the BC model as a function of the non-ideality interaction parameter, $\bar{\omega}$, for (a) temperatures and (b) densities. The symbols indicate calculations of liquid-gas critical points (MF - red closed circles, MC - green open circles), tricritical points (MF - red stars, MC - green stars), $\lambda$-end points (MF - blue triangles, MC - green triangles), and triple points between disordered liquid, ordered liquid, and vapor MF - purple diamonds). Black open circles, black curves, and red curves are the theory for multicritical behavior in a symmetric binary lattice model without the possibility of interconversion by Furman et al.~\cite{Furman_Global_1977}. In (a), the inset illustrates the multicritical behavior near the onset of tricriticality.}
    \label{Fig_CPvsWBar}
\end{figure}
As discussed in Section~\ref{Sec_Predictions}, the non-interconverting lattice model of Furman et al.~\cite{Furman_Global_1977} predicts tricritical and fourth-order critical phase behavior in the symmetric version of the binary mixture. A thorough comparison of different criticalities between the BC model and Furman et al.'s model are presented in Fig.~\ref{Fig_CPvsWBar} across all four archetypes. In type I and type IV behavior one may observe a bifurcation in the critical and $\lambda$-end point lines (red circles and blue triangle). In type III behavior, shown in detail in the inset on Fig.~\ref{Fig_CPvsWBar}a, there is an intersection of the critical, tricritical, and triple point regions. Lastly, the results of Furman et al. accurately predict the onset, termination, and behavior of tricriticality in type III behavior. While the meanfield results accurately follow the predictions of Furman et al., the MC simulation data for $Z_n=6$ only qualitatively agrees with these predictions.

\section{Conclusions}~\label{Sec_Conclusion}
In this work, we have introduced a concept of degenerate fluid polyamorphism in a single-component system where, in addition to evaporation and condensation terminated at the critical point (liquid-gas transitions), a line of second-order transitions between two different states of fluid can exist. Such polyamorphism is observed in $^4$He, where both first-order liquid-gas transitions and a line of second-order transition between normal and superfluid liquids are observed. Degenerate fluid polyamorphism can be caused by ``degenerated'' interconversion between two molecular states of fluid, in which the chemical potentials of individual (pure) states are always equal, meaning that the energy, entropy, and volume of interconversion are zero. The interconversion rate is assumed to be fast, without significant resistance. A simple example of resistance-free interconversion is the flipping of spins in the Ising model of ferromagnetism or the unrestricted interconversion of optically active chiral isomers.
	
We have studied analytically, in the meanfield approximation, and computationally, via 3D Monte Carlo simulations, a compressible, symmetric binary lattice-gas model with unrestricted interconversion of species. This system is thermodynamically equivalent to a single-component fluid, and is referred to as the degenerate BC model. It exhibits ``degenerate polyamorphism,'' in which both liquid-gas transitions and a line of second-order transitions (``lambda line'') between disordered liquid I and ordered liquid II exist, like that observed in $^4$He. Moreover, at certain combinations of the interaction parameters, lambda transitions may strongly interact with the liquid-gas equilibrium, generating symmetrical tricritical points. 
	
A distinct feature of degenerate fluid polyamorphism, which makes it fundamentally different from symmetric binary systems, even if the latter may exhibit similar phase diagrams at fixed equal amounts of species, is the change of the Gibbs phase rule. This change arises due to the interconversion of species which transforms the concentration from an independent thermodynamic variable to an equilibrium reaction coordinate that depends on temperature and pressure, thus making such a system thermodynamically equivalent to single-component fluids. 

Moreover, the phenomenon of phase amplification is characteristic of degenerate polyamorphic systems, and it causes a second-order transition line between fluids I and II to emerge. Fundamentally, phase amplification is associated with the symmetry of the nonconserved vector-like order parameter. This phenomenon does not occur in symmetric binary systems where the concentration is a conserved scalar property. In addition, we find that for large degrees of non-ideality of the system, the $\lambda$-line may intercept the gaseous branch of liquid-vapor coexistence, such that well below the liquid-gas critical point, an order-disorder transition between two gaseous states may occur. Indeed, the interception of the $\lambda$-line with the gaseous branch of liquid-vapor coexistence has been observed in the chiral model of interconverting enantiomers~\cite{Wang_Chiral_2022}.

There are a few important questions that could be a subject of future studies. First, it would be interesting to perform MC simulations of the BC model for a broad range of coordination numbers to investigate the crossover from fluctuation-induced criticality to meanfield criticality in the vicinity of the tricritical point. Second, a natural follow-up to this study would be to perform off-lattice molecular dynamics simulations of the BC model to investigate the effects of long-range interaction potentials on the emergence of the order-to-disorder transition in the gaseous phase for large degrees of non-ideality. Third, it would be interesting to investigate the chiral model~\cite{Latinwo_MolecModel_2016,Uralcan_Interconversion_2020,Petsev_Effect_2021,Wang_Chiral_2022} in a wider range of the model's ``biased parameter'' (physically equivalent to the nonideality parameter $\bar{\omega}$ in the BC model) and at different degrees of resistance to interconversion controlled by the model's ``rigidity parameter.''

\appendix
\section{Appendix: Emergence of Tricriticality in a Degenerate BC Model as Predicted by Landau Theory of Phase Transitions}\label{Sec_Appendix_Landau}

The application of the Landau theory describing the emergence of a tricritical point has a long history~\cite{Landau_Theory_1937,LL_Stat_Phys,Anisimov_Coupled_1981,anisimov_critical_1991,Roux_Sponge_1992,Wilding_Symmetrical_1998,Anisimov_Mesoscopic_2024}. In this Appendix, we qualitatively demonstrate the emergence of tricriticality in a degenerate BC model as predicted by the Landau expansion of the free energy in powers of two coupled order parameters. 

The Helmholtz free energy per unit volume, $f$, of a degenerated Ising-like fluid with interconversion of species, which possesses two order parameters, a vector-like $\psi$ and a scalar  $\varphi$, may be written in the form~\cite{Anisimov_Coupled_1981,Wilding_Symmetrical_1998,Anisimov_Mesoscopic_2024}
\begin{equation}
    f(T,\psi,\varphi) = f_\psi(T,\psi) + f_\varphi(T,\varphi) + f_\times(\psi,\varphi)
\end{equation}
where $f(T,\psi)$ is the free every of a degenerate incompressible system, in which the density is $\rho=1$, and the order parameter is defined as $\psi = 2x-1$, where $x$ is a fraction of interconversion; $f(T,\varphi)$, is the free energy of a degenerated compressible binary lattice with always equal contents and chemical potentials of pure species, in which the order parameter is $\varphi = 2\rho-1$; and $f_\times(\psi,\varphi)=\lambda\psi^2\varphi$ is a part of the free energy that determines a coupling between the order parameters, in which $\lambda$ is the coupling constant. Near the uncoupled second-order transition between two liquids, I and II and near the uncoupled liquid gas critical point, the Landau expansion in powers of the order parameters has the form
\begin{equation}\label{App_Eq_GenFreeEng}
    f(T,\psi,\varphi) = \frac{1}{2!}a_\psi\Delta\hat{T}_\psi\psi^2 + \frac{1}{4!}u_\psi\psi^4 + \frac{1}{6!}g_\psi\psi^6 + \frac{1}{2!}a_\varphi\Delta\hat{T}_\varphi\varphi^2 + \frac{1}{4!}u_\varphi\varphi^4 - h_\varphi\varphi+ \lambda\psi^2\varphi
\end{equation}
where $a_\psi$, $u_\psi$, $g_\psi$, $a_\varphi$, and $u_\varphi$ are the Landau-expansion coefficients, while $\Delta\hat{T}_\psi$ and $\Delta\hat{T}_\varphi$ are the reduced temperature distance to the corresponding critical temperatures, $T_\psi$ and $T_\varphi$, of the uncoupled phase transitions, $T_\psi=T_\lambda(\rho=1)$ and $T_\varphi=T_\text{c}(\bar{\omega}=0)$. In Eq.~(\ref{App_Eq_GenFreeEng}), $h_\varphi = a_\varphi\Delta\hat{T}_\varphi\varphi + (1/3!)u_\varphi\varphi^3$ is a ``field'', in the lowest approximation, linear to $\varphi$, conjugate to the scalar order parameter. For a lattice-gas model, $a_\psi=a_\varphi=1$ and $u_\psi=u_\varphi=2$. We note that the field conjugate to the vector-like order parameter for a degenerate interconversion is always zero.

Minimizing the free energy with respect to $\varphi$, $\partial f/\partial\varphi = 0$, one finds an effective scalar order parameter, $\tilde{\varphi}$, affected the coupling with $\psi$
\begin{equation}~\label{App_Eq_ConsOrderParamMinized}
    \tilde{\varphi} = \varphi -\lambda\psi^2\hat{\chi}_\varphi
\end{equation}
where $\varphi = h_\varphi\hat{\chi}$ is the uncoupled value of the scalar order parameter and $\hat{\chi}_\varphi = \partial\varphi/\partial h_\varphi = \Delta\hat{T}_\varphi+(1/3!)u_\varphi\varphi^2$ is the susceptibility, in the lowest approximation, for the uncoupled liquid-gas transition.

In the lowest-order approximation, the coupled $\lambda$-transition temperature decreases upon decrease of the density as $T_\lambda=T_\psi - 2\lambda\varphi$. For small values of $\varphi$, the density scales as $\psi^2$, as demonstrated in Fig.~\ref{Fig_comboX}.

Using Eq.~(\ref{App_Eq_ConsOrderParamMinized}), we may write free energy as a function of a single order parameter, $\psi$, 
\begin{equation}\label{App_Eq_FreeEn_Comb}
    f(T,\psi)=\frac{1}{2!}\Delta\hat{T}_\lambda\psi^2 + \frac{1}{4!}u_\lambda\psi^4 + \frac{1}{6!}g_\psi\psi^6
\end{equation}
in which the distance to the $\lambda$-transition is renormalized by the coupling of the order parameters with the coupling constant
\begin{equation}
    \lambda = -\frac{1}{2T_\lambda}\dv{T_\lambda}{\varphi}
\end{equation}
Since the $\lambda$-temperature at the densities close to $\rho=1$ is 
\begin{equation}
    T_\lambda=2\rho\bar{\omega}\left[1 - \frac{1}{T_\lambda} \dv{T_\lambda}{\varphi} \varphi\right]    
\end{equation}
(see Section~\ref{Sec_Methods_BCM}), the coupling constant, in the lowest-order approximation ($\rho\to 1$) is independent of $\bar{\omega}$, such that $\lambda\simeq 1/2$.

\begin{figure}[t!]
    \centering
    \includegraphics[width=0.7\linewidth]{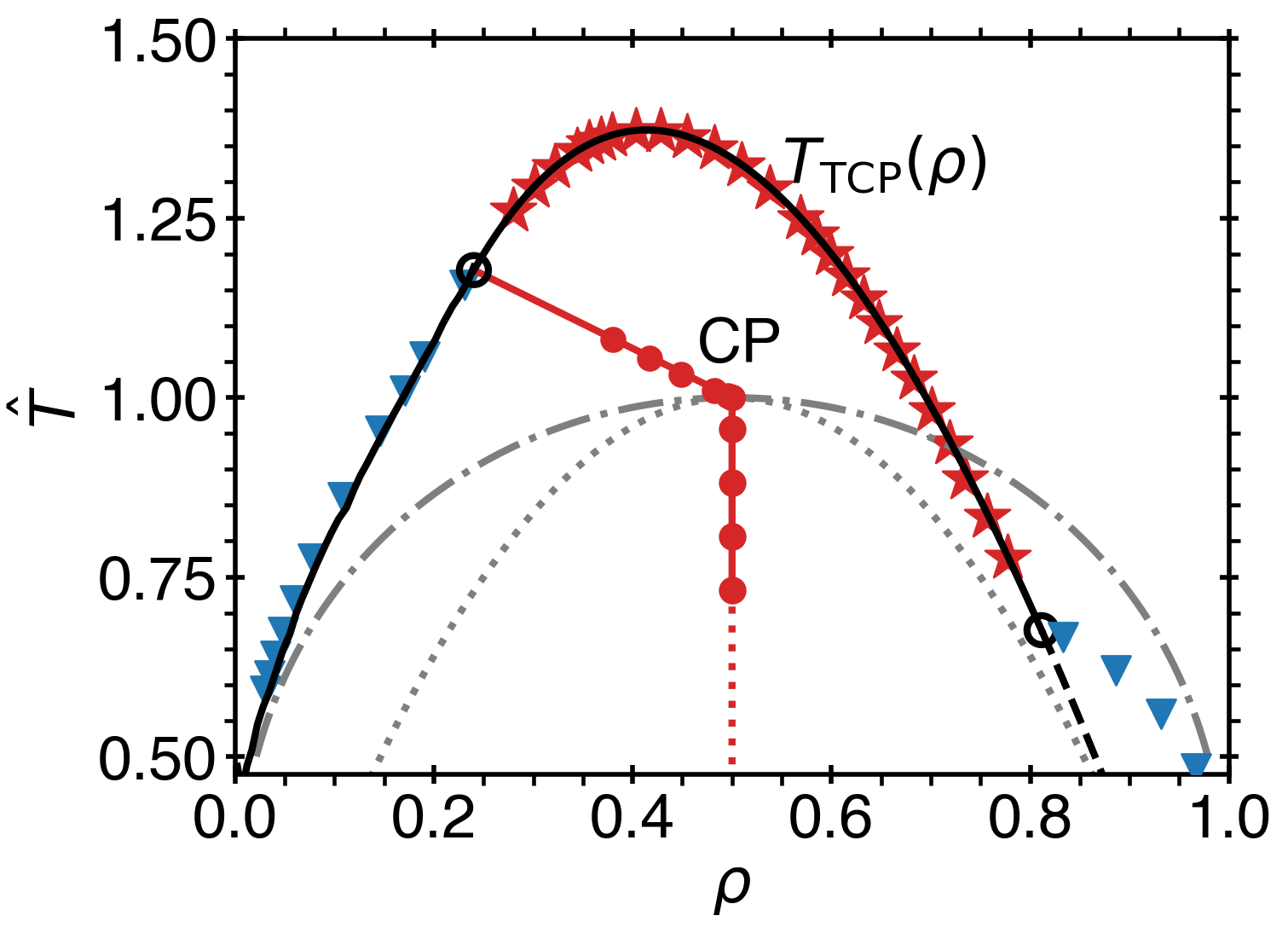}
    \caption{Meanfield prediction for the multicritical $T$-$\rho$ behavior in the degenerate BC model in the range of the non-ideality parameter from $\bar{\omega}=0.2$ (last triangle on the right) to $\bar{\omega}=10$ (first triangle on the left) with respect to the uncoupled liquid-gas transition ($\bar{\omega}=0$). The same data are also shown as a function of $\bar{\omega}$ in Fig.~\ref{Fig_CPvsWBar}. Gray curves represent the binodal (dash-dotted) and spinodal (dotted) of the uncoupled system. The symbols indicate the liquid-gas critical points (red closed circles), tricritical points (red stars), and $\lambda$-end points (blue triangles). The dotted red line indicates the continuation of the liquid-gas critical line into the metastable region. Open circles (at $\bar{\omega}=0.417$ on the right and $\bar{\omega}=2.45$ on the left) are the termination and emergence of tririticality and the solid lines are given by the analytic theory for multicritical behavior in a symmetric binary lattice model without interconversion~\cite{Furman_Global_1977}.}
    \label{Fig_tLVCPvsRhoLVCP}
\end{figure}

In addition, the square of the difference $(\tilde{\varphi}-\varphi)^2 \simeq \lambda^2\psi^4\hat{\chi}^{2}_\varphi$ contributes to the Landau expansion in powers of $\psi$, renormalizing the fourth-order coefficient, $u_\psi$, as 
\begin{equation}
    u_\psi \to u_\lambda \simeq u_\psi - 12\lambda^2\hat{\chi}_\varphi
\end{equation}
Beyond the linear approximation, the uncoupled susceptibility is a function of $T$ and $\varphi$ as
\begin{equation}
    \hat{\chi}^{-1}_\varphi = \pdv{\varphi}{h_\varphi} = \Delta\hat{T}_\varphi + \frac{1}{2}u_\varphi\varphi^2
\end{equation}
For the degenerated BC model, $u_\psi=2$, $\lambda=1/2$, hence $u_\psi\simeq 2\left\{1 - (3/2)/\left[\Delta\hat{T}_\varphi+\varphi^2\right]\right\}$. A locus defined by the condition
\begin{equation}
    \Delta\hat{T}_\varphi + \varphi^2 \simeq \frac{3}{2}
\end{equation}
Determines the boundary of the emergence of tricriticality, since the condition $\hat{\chi}^{-1}_\varphi = 0$ defines the liquid-gas spinodal. The boundary of the emergence of tricriticality is just a shifted spinodal of the uncoupled liquid-gas transition by a constant.  This prediction of the Landau theory is in qualitative agreement with the results demonstrated in Fig.~\ref{Fig_tLVCPvsRhoLVCP}.

The Landau expansion assumes that both $\Delta\hat{T}_\varphi$ and $\Delta\hat{T}_\psi$ as well as the coupling constant are small parameters. Therefore, the description given by the Landau expansion can only qualitatively be applied to the results presented for the degenerated BC model.

\begin{acknowledgement}
M.A.A. would like to acknowledge useful discussions with Pablo Debenedetti on the degenerate fluid-phase coexistence, caused by interconversion of enantiomers in the chiral model.
\end{acknowledgement}

\bibliography{ref}

\end{document}